\RequirePackage{fix-cm}

\documentclass{svjour3}

\smartqed  % flush right qed marks, e.g. at end of proof

\usepackage{graphicx}
\usepackage{xcolor}
\usepackage{hyperref}
\usepackage{diagbox}
\usepackage[utf8]{inputenc}
\usepackage{fancyhdr}
\usepackage{booktabs} 
\usepackage{amsmath} 
\usepackage{booktabs}
\usepackage{multirow}
\usepackage{siunitx}
\usepackage{tabularx}
\usepackage{array}
\usepackage[most]{tcolorbox}
\usepackage{pdfpages}

\title{Toward Inclusive AI-Driven Development: Exploring Gender Differences in Code Generation Tool Interactions}

\author{
Manaal Basha\textsuperscript{1}, 
Ivan Beschastnikh\textsuperscript{1}, \\
Cleidson R. B. de Souza\textsuperscript{2}, \\
Gema Rodriguez-Perez\textsuperscript{1}\\
\emph{\textsuperscript{1}University of British Columbia}\\
\emph{\textsuperscript{2}Federal University of Pará}
}
\institute{{\small \texttt{manaals@student.ubc.ca}, 
\texttt{bestchai@cs.ubc.ca},
\texttt{cleidson.desouza@acm.org},\\
\texttt{gema.rodriguezperez@ubc.ca}
}}

\date{May 2026}

\titlerunning{Gender Differences in Code Generation Tool Interactions}

\authorrunning{Basha et al.}

\begin{document}
%\includepdf[pages=1]{Authors_contact_.pdf}

\maketitle

\begin{abstract}
The increasing reliance on Code Generation Tools (CGTs), such as Claude Code and GitHub Copilot, is revamping programming workflows and raising critical questions about fairness and inclusivity in human-AI collaboration. While CGTs offer potential productivity enhancements, their effectiveness across diverse user groups have not been sufficiently investigated.  We hypothesized that developers' interactions with CGTs vary based on gender, influencing task outcomes and cognitive load, as prior research suggests that gender differences can affect technology use and cognitive processing. This study employed a mixed-subjects design with 39 participants, evenly divided by gender for a counterbalanced design. Participants completed two programming tasks of medium to high difficulty using two distinct treatments: only CGT assistance and only internet access. Task orders and conditions were counterbalanced to mitigate order effects. We collected cognitive load surveys, screen recordings, and task performance metrics such as completion time, code correctness, and CGT interaction behaviors. %Statistical analyses was conducted to identify statistically significant differences in CGT usage. 

Our results indicate no statistically significant gender differences in cognitive load or performance outcomes when using CGTs compared to Internet-based workflows. CGTs reduce intrinsic and extraneous cognitive load compared to Internet-based workflows, but the reduction was not statistically significantly. However, CGTs improved advanced code correctness. %We also find that cognitive load, particularly extraneous load, is associated with increased completion time. % We did not observe any statistically significant gender differences in this sample when using CGTs or the Internet to assist with coding. 
Our results suggest that CGTs can lower cognitive load and enhance performance on complex coding tasks without significantly affecting core correctness or completion time. These findings highlight how CGT usage can reduce cognitive burden and support more equitable programming experiences across users.

\end{abstract}


\section{Introduction}

Code Generation Tools (CGTs) based on Large-Language Models (LLMs), such as Windsurf\footnote{https://windsurf.com/} and GitHub Copilot\footnote{https://github.com/features/copilot}, are at the forefront of software development with millions of paid users~\cite{Ramel2024}. Studies on user perceptions of CGTs highlight several of their benefits, especially for novice users~\cite{penney2025outcomes,prather2024widening}. However, significant socio-technical challenges remain, including issues of trust~\cite{%VaithilingamZhangGlassman2022,
cheng2024would,wang2024Educators} and debugging complexity~\cite{Chasins2022}, which users often struggle with. Over-reliance on CGTs and distracting tool features~\cite{Becker2023} have also been identified as significant concerns. Additionally, there is a general fear that CGTs may introduce code vulnerabilities~\cite{Ahmad2022,kulkarni2024}. Despite these concerns, CGTs are designed to reduce cognitive load during certain tasks~\cite{Mendes2024,Glassman2020}.

Significant progress has been made in addressing these issues, with efforts underway to develop more transparent tool features~\cite{Patel_2024,tenney2024interactive}, improve tool accuracy through recent model enhancements~\cite{liu2024your}, and promote responsible AI practices by establishing and adhering to standardized guidelines~\cite{%hacker_2023,
constantinides2024rai,omar_2023}. Our work aligns with the fairness and transparency dimensions of the FATE\footnote{https://www.microsoft.com/en-us/research/theme/fate/} (Fairness, Accountability, Transparency, Ethics) framework by investigating potential biases in CGT usability, focusing on gender disparities in task performance and tool interactions, with the aim of informing more equitable tool design. While prior work has explored linguistic barriers in AI-assisted coding~\cite{buscemi2023comparative,Koyanagi2024,wang2022mconala}, \emph{gender-based} disparities remain under-explored. 

Addressing gender differences in CGT tools is particularly important as gender diversity is a key factor in ensuring that these tools serve a broad range of users effectively. The training data used by CGTs can reflect existing biases; for example, GPT-3 has been shown to exhibit biases due to the text it was trained on~\cite{gururangan2022whose}. In cases of language evaluation biases can surface in line with stereotypes~\cite{yamashita2025exploring}. Similarly, the code and comments used to train CGTs are predominantly sourced from a male-dominated industry~\cite{daniel2013effects}. If these models implicitly encode the preferences or mental models of a single demographic, they may inadvertently reinforce existing biases and create usability challenges for others groups. %which may lead these tools to unintentionally cater to a narrow demographic. This could reinforce existing biases and limit the tools' accessibility and usefulness for other groups. 

We hypothesize that gender differences exist in how CGTs %, trained on datasets predominantly created by male developers, 
are used. This study investigates the influence of gender on CGT usage not to generalize or reinforce divisions, but to reveal potential mismatches between tool design and user needs. 
Insights from our work can inform the development of CGTs that are more inclusive and responsive to the cognitive and behavioral diversity across developer populations, particularly in settings where team composition cannot change but the tools can be made more accommodating.
%
% Insights from our work can inform the development of CGTs that are more inclusive and responsive to the cognitive and behavioral diversity across developer populations. Particularly in settings where team composition cannot change, but the tools can be made more accommodating. 
% contributing to a more inclusive and equitable understanding of these transformative technologies.

To validate our hypothesis, we evaluate gender difference in CGTs in terms of cognitive load, task outcome, and CGT utilization as compared to the standard alternative of using the internet. This comparison is necessary to establish a benchmark for how gender differences manifest in a familiar and widely-used alternative to CGTs. Forums, documentation, and search engines are resources available on the internet that developers often rely on for programming support. By comparing gender differences in cognitive load and task outcomes, we can better understand whether CGTs offer unique challenges or advantages that differ from more traditional methods from the internet, and identify if they present gender-related disparities.%We aim to determine significant differences that can be improved upon to enhance the user adoption and experience with CGTs relative to gender. 

\textbf{Summary of Results and Contributions:} Our findings show that there were no statistically significant gender differences across any of the metrics for our sample. Further, participants using CGTs demonstrated improved advanced code correctness compared to Internet-based workflows, while also experiencing lower intrinsic and extraneous cognitive load, although these reductions were not statistically significant. We further observed that higher extraneous cognitive load was associated with increased task completion time. While no statistically significant gender differences were found in overall performance outcomes, notable differences emerged in interaction behaviors and workflow strategies.

We show that although CGTs reduce intrinsic and extraneous cognitive load, these reductions do not consistently translate into improvements in correctness or completion time, challenging common assumptions in prior work. Our results further reveal a shift from incremental programming to generate-and-test workflows, helping explain this disconnect between perceived efficiency and objective performance. We also identify an efficiency–engagement trade-off, where reduced effort is accompanied by lower engagement with code and increased reliance on generated outputs. Finally, we demonstrate that while CGTs yield comparable outcomes across genders, differences in interaction patterns highlight the need for process-based evaluations of fairness in AI-assisted development.

\section{Related Work}

%\subsection{Integration of CGTs}

CGTs are now integrated into most modern Integrated Development Environments (IDEs), offering features like built-in auto-completion \footnote{As seen in IntelliSense for VS Code and IntelliJ.}. Advanced paid versions, such as GitHub Copilot and Tabnine, have further normalized CGT usage in software development, aiming to enhance productivity. Empirical studies of practitioner experiences with tools like Copilot reveal persistent operational, compatibility, and reliability challenges \cite{zhou2025exploring}. However, this widespread adoption also raises questions concerning Fairness, Accountability, Transparency, and Ethics (FATE).

\subsection{Gender and Bias in Software Engineering}

Terrell et al.~\cite{terrell2017gender} highlighted the presence of implicit bias in technical domains by showing that women’s contributions to open-source projects are accepted at higher rates when their gender is not identifiable, but at significantly lower rates when it is known. These findings suggest the existence of systemic barriers that are not immediately visible or addressed. The authors concluded that more efforts are needed to create a more inclusive environment for diverse user groups. More recently, Treude and Hata~\cite{Treude2023} used a back-translation approach and identified a significant disparity in the gendered pronoun associations with different software development tasks. For instance, requirements activities were associated 6\% of the time with men, while testing activities were 100\% associated with men.

To provide a structured approach to evaluate gender inclusivity, Burnett et al. introduced the GenderMag framework~\cite{burnett2016gendermag}. They identified five cognitive facets known to differ by gender: attitude toward risk, self-efficacy, information processing style, motivations, and learning preferences. For example, women tend to exhibit lower self-efficacy and higher risk aversion when engaging with technology~\cite{beckwith2005common,burnett2016gendermag}, which may influence how they interact with CGTs. 

These factors have been shown to significantly impact decision-making and interactions with technology. For instance, risk-averse individuals may hesitate to explore or adopt new tools like CGTs, while those who are more risk-tolerant are likely to experiment and engage more freely. This pattern was observed by Durndell and Haag~\cite{durndell2002computer} in a study with 150 participants during the early days of internet adoption. The study assessed participants’ experiences using various tests and revealed notable gender differences. Half of the participants were female, and compared to males, they reported spending less time using the internet, holding less positive attitudes toward it, experiencing greater computer-related anxiety, and demonstrating lower computer self-efficacy.

Additionally, the automation inherent in CGTs can influence users' learning styles. For instance, CGTs often suggest code snippets, making them particularly useful to users who prefer exploratory or tinkering-based learning approaches. These users may experiment with the suggestions, using trial and error to refine the code and resolve bugs, ultimately aiming to produce correct and functional code. However, research has shown that women tend to exhibit less interest and confidence in tinkering behaviors~\cite{burnett2010gender}, which are critical for maximizing the potential of CGTs.

This distinction aligns with findings from Barke et al. ~\cite{barke2023grounded} who identified two distinct user states in their study of CGT interaction trends. Users in an \emph{exploratory} state engaged in tinkering, experimenting with tool features, and using trial-and-error methods to navigate coding challenges. In contrast, users in an \emph{acceleration} state leveraged CGTs to enhance their performance, applying the tools efficiently because they already knew what to code and how to implement it. These differing approaches highlight how user traits and learning preferences can shape effective use of CGTs.

\subsection{Developer Expertise and Cognitive Load}

Beyond gender-specific traits, the level of developer experience plays a critical role in the adoption and efficacy of CGTs. Ferino et al.~\cite{ferino2025novice} recently conducted a systematic literature review on novice developers’ perspectives, highlighting that while LLM-based tools lower the barrier to entry for beginners, they also introduce significant risks related to over-reliance and ``blind trust.'' Their synthesis reveals that novices often struggle with the validation of AI-generated code, a challenge compounded by the exploratory or ``tinkering'' states identified by Barke et al.~\cite{barke2023grounded}. Specifically, while experts may use CGTs for acceleration, novices frequently remain in an exploratory loop where the inability to distinguish between correct and ``hallucinated'' code can increase cognitive load. This suggests that without high self-efficacy—which often varies by gender~\cite{burnett2016gendermag}, novices are particularly vulnerable to the pitfalls of automated suggestions, which can hinder their long-term learning trajectory.

Gender differences in cognitive load have also been observed in various learning contexts, shedding light on how technology design and interaction types can affect user experiences. For instance, Chen et al.~\cite{chen2021gender} studied the cognitive load of school-aged children using learning robots and found that boys experienced significantly lower cognitive load than girls, despite no significant difference in their learning performance. The boys appeared to benefit more from the robots, which the authors attributed to the robot interface design favoring visual and spatial elements. These elements have been shown to align better with male cognitive strengths, while females tend to benefit more from auditory components~\cite{pauls2013gender}. Such insights highlight the importance of examining how CGT interfaces might inadvertently cater to certain cognitive strengths, potentially influencing their accessibility and effectiveness for different genders.

%There is a clear gap in research regarding how gender influences the use of CGTs. While some studies have examined gender differences in broader technological contexts, the specific ways in which these differences affect the adoption, interaction patterns, and effectiveness of CGTs remain unexplored. Our research aims to fill this gap by investigating whether 

Building on prior studies that highlight gender-based differences in technology use and adoption~\cite{burnett2016gendermag,durndell2002computer}, we hypothesize that similar disparities exist in how users interact with CGTs. Research has shown gendered differences in cognitive load across learning environments, particularly when interfaces align with certain sensory modalities~\cite{chen2021gender,pauls2013gender}, motivating our investigation into whether CGTs similarly create unequal cognitive demands (RQ1). Further, women have been found to engage less in exploratory or tinkering-based programming behaviors~\cite{burnett2010gender}, which are linked to CGT effectiveness~\cite{barke2023grounded}, raising questions about gender-based differences in both task outcomes (RQ2) and tool utilization (RQ3). This study ultimately aims to provide evidence that can guide the design of more inclusive and equitable CGTs, ensuring they better serve a diverse user base.

\section{Research Questions and Hypotheses}

Prior research has reported demographic differences in technology interaction, including variations in self-efficacy, exploratory behavior, cognitive load, and software development practices~\cite{beckwith2005common,burnett2010gender,burnett2016gendermag,durndell2002computer}. However, it remains unclear whether these differences persist in AI-assisted programming environments, where CGTs introduce new interaction paradigms through conversational prompting, automated code synthesis, and iterative code refinement. 

This study aims to evaluate gender associated differences in the use of CGTs during programming tasks as compared to traditional internet assisted workflows. Specifically, we focus on cognitive load, task performance, and interaction patterns as these dimensions collectively capture how developers experience, complete, and engage with programming tasks when using CGTs. Together, they provide insight into both the perceived and behavioral effects of AI-assisted development, enabling a more comprehensive evaluation. Based on the prior studies described above, we defined the following research questions to guide this investigation:

\textbf{RQ1: Are there gender differences in the cognitive load (intrinsic, extraneous, and germane load scores) experienced by participants when using CGTs to complete programming tasks?}
Cognitive load can affect how developers interact with CGTs and their overall performance. This question investigates whether gender influences cognitive load during the use of CGTs. %Based on the results seen by Burnett et al.~\cite{burnett2016gendermag} and Chen et al.~\cite{chen2021gender} we hypothesize that there will be a statistically significant difference in each type of cognitive load between participants of different genders when using CGTs. 

\textbf{H1:} Based on prior literature on technology interaction and cognitive load~\cite{burnett2016gendermag,chen2021gender}, we hypothesize that there will be a statistically significant difference in each type of cognitive load between participants of different genders when using CGTs. 

%This hypothesis is motivated by prior work that identifies gender-related differences in software development tools and cognitive processing styles~\cite{burnett2016gendermag, chen2021gender}.

\textbf{RQ2: How do CGT outcomes differ between participants of different genders?} %(correctness, functionality, proportion of final code that is CGT-generated, suggestion acceptance rate (SAR), generated code that was modified after being accepted, and completion time)
This question evaluates whether gender impacts task outcomes (i.e., correctness and completion time) when using CGTs. 

%Prior studies on programming behavior and AI-assisted development suggest that interaction strategies, verification practices, and exploratory workflows may influence programming performance outcomes~\cite{barke2023grounded}

\textbf{H2:} Drawing on prior research on programming interaction behavior and AI-assisted development workflows~\cite{barke2023grounded}, we hypothesize that there will be statistically significant differences in task outcomes between participants of different genders when using CGTs. %This hypothesis is informed by previous studies that shown that users often engage in tinkering while using CGT~\cite{barke2023grounded}. 
%, such as the correctness, functionality, and efficiency of the completed tasks. 
%Based on the work of Beckwith et al.~\cite{beckwith2005common}, Durndell et al~\cite{durndell2002computer}, and Burnett et al.~\cite{burnett2010gender}, we hypothesize that there will be statistically significant differences in task outcomes %(correctness, functionality, proportion of final code that is CGT-generated, suggestion acceptance rate (SAR), generated code that was modified after being accepted, and completion time) 
%between participants of different genders when using CGTs. Further influenced by Pauls et al.~\cite{pauls2013gender}, we also hypothesize there will be a statistically significant difference in task outcomes relative to using a CGT versus the baseline approach.

\textbf{RQ3: How does CGT utilization differ between participants of different genders?} %(correctness, functionality, proportion of final code that is CGT-generated, suggestion acceptance rate (SAR), generated code that was modified after being accepted, and completion time)
This question evaluates whether gender impacts tool utilization (i.e., proportion of final code that is CGT-generated, suggestion acceptance rate (SAR), generated code that was modified after being accepted, prompt generation count) when using CGTs.

%Prior literature suggests that users may differ in how they engage with software tools, including exploratory behavior, feature usage, confidence in generated outputs, and willingness to modify or verify automated suggestions~\cite{beckwith2005common,durndell2002computer,burnett2010gender}. In the context of CGTs, such differences may manifest through prompting behavior, suggestion acceptance, reliance on generated code, and modification of generated outputs.

\textbf{H3:} Based on prior work on technology interaction and programming behavior~\cite{barke2023grounded,beckwith2005common,burnett2010gender,durndell2002computer}, we hypothesize that there will be statistically significant differences in tool utilization between participants of different genders when using CGTs. %This hypothesis is informed by previous research~\cite{beckwith2005common,durndell2002computer,burnett2010gender, barke2023grounded}.

%While CGTs are often positioned as productivity-enhancing development tools, it remains unclear whether they fundamentally alter programming workflows relative to traditional internet-assisted approaches. Comparing these workflows helps contextualize observed interaction behaviors and cognitive load patterns.

\textbf{H4:} Based on prior work on interaction models and information access patterns ~\cite{pauls2013gender}. We hypothesize that there will be a statistically significant difference in tool utilization when using a CGT compared to the internet approach. %This %expectation hypothesis is further supported by the findings in Pauls' et al work~\cite{pauls2013gender}.
%, such as the correctness, functionality, and efficiency of the completed tasks. 
%Based on the work of Beckwith et al.~\cite{beckwith2005common}, Durndell et al~\cite{durndell2002computer}, and Burnett et al.~\cite{burnett2010gender}, we hypothesize that there will be statistically significant differences in tool utilization %(correctness, functionality, proportion of final code that is CGT-generated, suggestion acceptance rate (SAR), generated code that was modified after being accepted, and completion time) 
%between participants of different genders when using CGTs. Further influenced by Pauls et al.~\cite{pauls2013gender}, we also hypothesize there will be a statistically significant difference in tool utilization relative to using a CGT versus the baseline approach.

%\textbf{RQ3: Are there differences in the perceived usefulness of CGTs compared to the internet between participants of different genders in helping them complete programming tasks?}
%Perceptions of usefulness may vary by gender and impact the adoption and use of CGTs. This question assesses whether gender or condition influences how participants perceive the value of CGTs in completing their tasks. We hypothesize that there will be no significant differences in the perceived usefulness of CGTs between participants of different genders.

\section{Methodology}

This section outlines the experimental design, participant recruitment, task selection, counterbalancing strategies \cite{gergle2014experimental}, and metrics used in the study to explore gender differences in utilizing CGTs. %The goal is to investigate how the cognitive load and task performance differ between genders when using CGTs like Windsurf and compare it with traditional programming practices using the internet.

\subsection{Participants:}

To ensure meaningful results from our sample size, we conducted a power analysis following recommendations from Brysbaert~\cite{brysbaert2019many}. Assuming a medium effect size ($d \approx 0.4$), a significance threshold of $\alpha = 0.05$, and a desired statistical power of 0.95, the analysis indicated that a minimum of 16 participants would be required for the planned mixed-design comparisons. The selected effect size ($d \approx 0.4$) reflects a theoretically meaningful and practically interpretable medium-sized effect commonly used in behavioral research power analyses~\cite{brysbaert2019many}. However, smaller demographic effects may still exist and remain undetectable within studies of this scale.

A sample size of at least 16 participants was required for this mixed-subjects comparison. We recruited a total of 39 participants through university posters and social media advertisements. The final sample included approximately balanced gender representation, consisting of 19 females and 20 males. Participants were required to be graduate students or senior undergraduate students in computer science or closely related fields, have prior programming experience, with a focus on Python, and be adept with basic concepts of coding. Participants with no prior experience in Python or CGT usage were excluded to ensure that observed effects were less likely to be driven by insufficient programming proficiency or unfamiliarity with CGTs. The demographic and experience characteristics of the participants are summarized in Table \ref{tab:participants}.

\begin{table}[ht!]
\centering
\caption{Participant Demographics and Background Characteristics}
\label{tab:participants}
\begin{tabular}{lcc}
\toprule
 & \textbf{Count} & \textbf{Percentage (\%)} \\
\midrule

\textbf{Gender} & & \\
Male & 20 & 53.8 \\
Female & 19 & 46.2 \\
\midrule

\textbf{Age} & & \\
18--24 & 18 & 46.2 \\
25--34 & 21 & 53.8 \\
\midrule

\textbf{Education Level} & & \\
Undergraduate (Year 4+) & 25 & 64.1 \\
Graduate (Masters) & 8 & 20.5 \\
PhD & 6 & 15.4 \\
\midrule

\textbf{Python Familiarity} & & \\
$\sim$1 year & 7 & 17.9 \\
2--3 years & 21 & 53.8 \\
4+ years & 11 & 28.2 \\
\midrule

\textbf{CGT Usage Frequency} & & \\
Always & 16 & 41.0 \\
Most of the time & 13 & 33.3 \\
Sometimes & 9 & 23.1 \\
Rarely & 1 & 2.6 \\
\midrule

\textbf{Ethnicity} & & \\
South Asian & 22 & 56.4 \\
Asian & 7 & 17.9 \\
African-American & 3 & 7.7 \\
East Asian (China) & 2 & 5.1 \\
Hispanic or Latino & 2 & 5.1 \\
White/Caucasian & 2 & 5.1 \\
West Asian (Iran) & 1 & 2.6 \\
\bottomrule
\end{tabular}
\end{table}

%Participant-level behavioral characteristics reveal substantial heterogeneity in task performance and interaction patterns beyond the demographic distributions summarized in Table~\ref{tab:participants}. 

Participants completed a pre-study screening and demographic questionnaire to confirm eligibility and collect background information including age, programming experience, CGT usage frequency, and self-identified gender.% A post-study survey was additionally administered to capture subjective experiences and qualitative reflections that contextualized the quantitative findings.

This study focused on binary gender comparisons because no participants self-identified outside the categories of male or female during recruitment. Consequently, the findings should not be interpreted as representative of broader gender-diverse populations, which remain important directions for future work.

\subsection{Controlled Experimental Design}

Our study employed a mixed-subjects design, incorporating both within-subjects (task type and CGT condition) and between-subjects (gender) variables. Each participant completed two Python programming tasks: one using a CGT and one using the internet. Windsurf\footnote{Experiments were conducted using a Version 1.9 base model SWE-check release.}, a free and accessible CGT that integrates with VS Code and supports both auto-suggestions and chat-based prompting, was used due to its availability, cost-effectiveness, and ease of integration into developer workflows.

Participants could use anything on the internet that was not AI-related such as StackOverflow or Reddit, but could not use Google's AI summarizer for example.

Participants used their own laptops in a private lab space or remotely based on the participant's comfort, ensuring familiarity with their development environment and reducing variability from unfamiliar hardware. This within-subjects structure reduced individual difference effects, as each participant acted as their own control. Each participant recorded their screen, which was used to assist with and confirm metrics discussed in the next section.

Task 1 and Task 2 were designed to reflect realistic Python programming tasks that participants may encounter in practice while remaining comparable in overall effort and difficulty. Task 1 required participants to develop a web scraping script that extracted and structured upcoming movie event data into JSON format using libraries such as \texttt{requests} and \texttt{BeautifulSoup}, while Task 2 required participants to implement a directory monitoring system that tracked file system events and logged changes using \texttt{watchdog}. To ensure fairness across conditions, we iteratively refined both tasks to have similar lines of code (LOC) requirements and comparable cyclomatic complexity (CC) scores. Further, in our pilot participant feedback survey, participants rated the tasks as similar in difficulty. Since perceptions of medium-hard difficulty can vary depending on prior programming experience, we used LOC and CC to provide a more objective basis for balancing the tasks. The tasks were also intentionally designed around problem types that current CGTs could assist with but could not reliably solve end-to-end without participant involvement. This ensured participants needed to actively engage with the generated code through prompting, debugging, modification, and verification rather than simply copying complete solutions. Full task descriptions, templates, and materials are available in the replication package \cite{inclusiveai}.

To reduce order effects and increase internal validity, the order in which tasks and conditions were presented was counterbalanced. Participants were randomly assigned to one of two counterbalanced groups as seen in Table \ref{tab:hci-contingency}.

\begin{table}[ht!]
\centering
\small
\caption{Counterbalanced Task and CGT Conditions}
\label{tab:hci-contingency}
\begin{tabular}{|p{2.4cm}|p{2.5cm}|p{2.8cm}|}
\hline
\backslashbox{\textbf{  Tasks}}{\textbf{ CGT }} 
& \textbf{Starts With CGT} 
& \textbf{Starts Without CGT} \\
\hline
\textbf{Task 1 $\rightarrow$ Task 2} 
& Group A1 (With $\rightarrow$ Without) 
& Group B1 (Without $\rightarrow$ With) \\
\hline
\textbf{Task 2 $\rightarrow$ Task 1} 
& Group A2 (With $\rightarrow$ Without) 
& Group B2 (Without $\rightarrow$ With) \\
\hline
\end{tabular}
\end{table}

%By randomly assigning participants to one of these groups, we aimed to control for the effects of task order, ensuring that both conditions (CGT vs. internet) were presented to participants in a randomized sequence.

\subsection{Feature Selection and Statistical Modelling}

To evaluate the effects of gender and CGT usage on cognitive load, task performance, and tool utilization, we adopted a semiparametric modeling approach that accommodates the varied data types in our study. Semiparametric models combine parametric components (e.g., regression structure over predictors such as gender, task, and tool condition) with non-parametric or weak distributional assumptions, enabling flexible modeling without imposing strict constraints such as normality \cite{ruppert2003semiparametric}. This is particularly suitable given the heterogeneous nature of our outcome variables. 
Specifically, we leveraged ordinal regression models \cite{agresti2010analysis}, including the Proportional Odds Model (POM), for ordinal outcomes, as well as generalized linear (mixed) models tailored to outcome distributions, including linear mixed-effects models for continuous outcomes, negative binomial models for count data, and beta regression \cite{ferrari2004beta} for proportional outcomes.  

Our outcome variables $Y$ included: ordinal (ICL, ECL, GCL on a 9-point scale), count (prompt generation count), continuous (task time in seconds), and proportion (suggestion acceptance rate, CGT-generated code ratio, code modification rate, and correctness ratios) data types.

Ordinal response variables were analyzed using cumulative link models (proportional odds models, POMs), including mixed-effects extensions (CLMMs) where appropriate, to model ordered outcomes while accounting for repeated measures across participants, tasks, and conditions. These models estimate the effect of covariates (e.g., gender, task, tool condition) on the cumulative probability of higher cognitive load levels and do not assume normality, making them robust to extreme values.

Continuous outcomes (e.g., task completion time) were analyzed using linear mixed-effects models with participant-level random intercepts to account for repeated measures. Count data (e.g., prompt counts) were modeled using Poisson or negative binomial generalized linear mixed-effects models depending on dispersion diagnostics. Proportion outcomes (e.g., suggestion acceptance rate, correctness ratios) were modeled using beta regression after scaling to the open interval (0,1), appropriate for bounded continuous outcomes.

%Ordinal response variables were analyzed using POM, because it does not assume normality and is robust to extreme values. POM is particularly appropriate for Likert-type outcomes and allowed us to estimate how covariates (e.g., gender, task, tool condition) influenced cognitive load.

%Continuous outcomes (e.g., task completion time) were analyzed using robust linear regression to accommodate skewness and reduce sensitivity to non-normal errors. For count data (e.g., prompt count), we used Poisson or negative binomial regression, based on dispersion tests. Proportion outcomes (e.g., suggestion acceptance rate, code correctness) were scaled to (0,1) interval and modeled using beta regression.

To account for repeated measures (e.g., multiple tasks per participant), we included random intercepts for participant ID where appropriate, forming mixed-effects. This modeling framework (i.e., ordinal, linear mixed-effects, count, and beta regression) offers a unified approach to handling multiple outcome types, improving the interpretability and comparability of results while avoiding strong distributional assumptions. This strategy enabled a consistent and statistically robust framework for hypothesis testing across all dependent variables.

For measuring cognitive load, a 15 question survey utilizing a 9-point Likert-type scale ranging from 1 (not at all applicable) to 9 (fully applicable) was selected based on a meta-analysis by Krieglstein et al. \cite{krieglstein2022systematic} which found that the use of this scale in the existing cognitive load questionnaire was linked to higher reliability values. The questionnaire has also been shown to have strong internal consistency (e.g., Cronbach's $\alpha > .80$) and construct validity across multiple software and learning environments. We allocated 5 survey questions for each cognitive load type. %We allocated 5 survey questions for each cognitive load type, and based these on the recommended questionnaire found in~\cite{krieglstein2022systematic}. 
Each participant completed the full survey after each task to better understand the relationship between cognitive load and CGT usage. All surveys can be accessed via our replication package~\cite{inclusiveai}.

Post hoc analyses with multiplicity corrections were conducted as a robustness check to assess potential Type I error inflation, and these adjustments did not materially change any statistical conclusions or reveal additional significant effects.

% By randomly assigning participants to one of these groups, we aimed to control for the effects of task order, ensuring that both conditions (CGT vs. internet) were presented to participants in a randomized sequence.

\subsubsection{RQ1: Cognitive Load Measurement:}

To evaluate cognitive load, participants completed a survey after each task. These questions measured three different types of cognitive load~\cite{krieglstein2023development}:

\begin{itemize}
    \item \textbf{Intrinsic Load (ICL):} Cognitive effort due to task complexity; %The cognitive effort required due to the complexity of the task itself.
    \item \textbf{Extraneous Load (ECL):} Effort imposed by the environment or interface (e.g., distractions, poorly designed tool interfaces); and
    \item \textbf{Germane Load (GCL):} Effort involved in creating and maintaining schemas or mental models while solving the task.
\end{itemize}

These cognitive load scores were measured using an optimized questionnaire from Krieglstein et al.~\cite{krieglstein2023development}. %after each task and used to assess differences in cognitive load across gender, particularly in relation to the use of CGTs and the use of the internet. 
Alternative workload measures, such as the NASA Task Load Index (NASA-TLX), were considered for this study. However, prior research suggests that while the NASA-TLX is effective for measuring overall perceived workload, it may be less sensitive in distinguishing nuanced sources of cognitive effort and individual differences in cognitively demanding tasks~\cite{mckendrick2018deeper}. Although modified versions of the NASA-TLX have been proposed, such adaptations require additional validation to ensure reliability and construct validity~\cite{hart2006nasa}. Accordingly, this study adopts the multidimensional cognitive load questionnaire validated by Krieglstein et al.~\cite{krieglstein2023development}, which was specifically designed to reliably differentiate intrinsic, extraneous, and germane cognitive load. Differentiating these cognitive load dimensions is important in the context of CGT-supported programming, as it allows us to better isolate whether observed effects are associated with task complexity, workflow overhead, or productive cognitive engagement during problem solving.
%Alternative measures like the NASA Task Load Index (TLX) questionnaire were considered, but based on a construct validity study \cite{mckendrick2018deeper} with ten adults, researchers found that the questionnaire lacked sensitivity to personal capacities or task demands suggesting that the measure has limited utility. The authors advise that the approach is best when task demands are clearly perceptible. However, NASA advised that redefining the existing generic TLX metric labels to a relative situation can be a great strategy; however, these modified approaches need to be validated to ensure their reliability and sensitivity \cite{hart2006nasa}. Krieglstein et al. \cite{krieglstein2023development} reviewed and validated the cognitive load questionnaire used in this study by using 54 participants. The authors confirmed that the different types of cognitive load can be measured in a reliable and valid manner with the proposed questionnaire. The reason we differentiate the sources of cognitive load in this study is to better isolate the sources of the effects.

\subsubsection{RQ2: Task Outcome Metrics }

For RQ2 and RQ3 metrics we used CodeWatcher~\cite{basha2025codewatcher}, an open-source VS Code extension designed to capture fine-grained developer interactions during programming tasks. By logging timestamped behavioral events such as edits, focus changes, copy-paste actions, and CGT interactions, it enables objective analysis of developer workflows and AI-assisted programming behaviors beyond self-reported measures. We analyzed the following performance metrics using screen recordings and the CodeWatcher extension data:

\begin{itemize}

    \item \textbf{Task Completion Time:} Time taken to complete each task. This was measured from the moment the task started until the participant submitted their solution. This was calculated using the screen recordings.
    
    \item \textbf{Core Code Correctness:} Core correctness focuses solely on whether the code meets the essential requirements of the task for predefined, straightforward inputs. The inputs and expected outputs were the same as those provided in the user task instructions as examples. It was assessed as a percentage of test cases successfully passed. If the code produced the expected output for all specified unit test inputs, it was deemed correct. This ensures a baseline evaluation of the core logic, regardless of how well the code handles variations or additional nuances. If the code failed to run then it was given a correctness score of zero.
    
    \item \textbf{Advanced Code Correctness:} Advanced correctness refers to how comprehensively the submitted code satisfies the full range of task requirements, including the handling of edge cases and non-trivial input conditions beyond the core problem requirements. Advanced correctness was objectively evaluated by executing participant submissions against an extensive suite of automated test cases consisting of both standard scenarios and edge conditions, such as unusual input ranges, empty or null values, and high-volume data inputs. Scores were computed as the proportion of test cases successfully passed, reflecting the robustness and completeness of the submitted solution. Submissions that failed to execute were assigned a score of zero.
\end{itemize}

To examine how cognitive load relates to task outcomes, we fitted linear and generalized linear mixed-effects models with \textbf{completion time}, \textbf{core correctness}, and \textbf{advanced correctness} as dependent variables. Predictors included intrinsic (ICL), extraneous (ECL), and germane (GCL) cognitive load, tool condition (CGT vs Internet), task, and gender, with participant ID as a random intercept.

\subsubsection{RQ3: CGT Usage Metrics}
For CGT utilization, we used CodeWatcher~\cite{basha2025codewatcher} and participant recordings to collect the following metrics:

\begin{itemize}
    \item \textbf{Prompt Generation Count:} The number of times the CGT was prompted to generate code within the IDE chat window. This measured how actively participants relied on the CGT to produce solutions. This is different than the auto-suggestion feature that recommends code as the user types. This was measured by utilizing the participants' chat logs downloaded from the chat after the task was completed. A script was utilized that counted each time the user submitted a prompt from the logs.
    
    \item \textbf{Suggestion Acceptance Rate (SAR):} The percentage of generated code suggestions accepted by the participants, indicating how often they found the CGT's suggestions useful or usable. This was measured using CodeWatcher data and the recordings.
    \[
     \text{SAR} = \frac{\text{Num. of Accepted Code Suggestions}}{\text{Total Num. of Generated Code Suggestions}} \times 100
      \]
   
    \item \textbf{Modification Rate:} How often participants immediately modified the code suggested by the CGT that was accepted by the participant (e.g., changing variable names). This was measured using CodeWatcher data.
    
    \item \textbf{Proportion of Final Code Generated by CGT:} The percentage of code in the participant's final submission that was directly generated by the CGT compared to code written by the participants themselves. More specifically, the ratio of LOC that are AI written and the total LOC. This was measured using CodeWatcher data.

\end{itemize}

%These metrics were compared across gender to determine if there were any significant differences in how participants of different genders interact with CGTs.

CodeWatcher's performance was also randomly validated through manual confirmation using participant screen recordings.

\subsection{Ethical Considerations}

Ethics approval was granted for this study as H25-00502. We ensured informed consent, confidentiality via anonymization, and post-study debriefing. Participants received \$30 cash upon completion of our study.

\begin{itemize}
    \item \textbf{Informed Consent:} Participants were fully informed about the study’s objectives and procedures, and written consent was obtained before data collection began. Participants could also end their involvement in the study at any time without penalty.
    \item \textbf{Confidentiality:} Participants' personal information and data were kept confidential, and all recordings were anonymized.
    \item \textbf{Debriefing:} At the end of the study, participants were debriefed about the research objectives and results, and any questions were answered.
    \item \textbf{Compensation:} Participants received \$30 cash after their completion of the study.
\end{itemize}

\subsection{Threats to Validity}

\textbf{Internal Validity:} To ensure that our observed effects were due to changes in our independent variables and not from external factors, we used a within-subjects design. This design controls for individual differences and improves comparability. Further, prior programming experience was accounted for to ensure that participants did not lack a computing background, which can affect their overall task performance. All participants completed the same task under the same conditions to reduce variability in problem difficulty or environmental factors. To mitigate order effects, we counterbalanced task order across participants and included breaks in the session to reduce fatigue. In addition, we acknowledge the potential for subjective bias for survey submissions and complement this with performance-based metrics. Gender perception of task difficulty can lead to differences in perceived difficulty and could influence task performance. We account for this by measuring task ICL, which helps determine if there is any effect of gender on task difficulty. Participants were informed that screen recordings were used strictly for analyzing CGT interaction patterns, not performance evaluation, to reduce observation bias.

\textbf{External Validity:} To improve our findings' generalization to real-world scenarios, we ensured that tasks were designed to reflect real-world software development challenges, making findings more applicable to professional developers. Further, we selected participants in their final year of undergraduate study and above as they are more likely to have exposure to internships and co-ops, ensuring industry-relevant skills. However, as our sample consists of students, our findings may not fully capture industry professional behaviors, or can miss smaller demographic effects. Additionally, this study focuses on a specific CGT; results may not generalize to all CGT tools. Although this study uses Windsurf as the representative CGT, its prompt-based, inline suggestion paradigm is shared across tools like GitHub Copilot. As such, we expect our findings to generalize to similar tools, though future work should validate these results across different CGT interfaces and programming environments. Further, regional and cultural factors may influence results, and future studies should consider an even larger, more diverse participant pool.

\textbf{Construct Validity:} The metrics used in this study, such as task completion time and task ICL, aim to allow us to better understand the cognitive effort and difficulty, but they may not fully capture subjective workload or user experience. Additionally, differences in familiarity with VS Code or Windsurf may introduce unintended variability in performance.

\section{Deviations from the Registered Report Protocol}

This study primarily followed the approved Phase 1 Registered Report protocol \cite{basha2025toward}; however, some deviations were made to improve the completeness and reliability of the analysis.

First, although our initial best case recruitment target was 54 participants, we recruited a total of 39 participants due to recruitment limitations. We increased the initial incentive from \$25 to \$30 to increase participation. This sample size remained above the minimum required sample determined by our power analysis. Using a significance threshold of $\alpha = 0.05$ and 95\% statistical power, the analysis indicated that at least 16 participants were required for the planned mixed-subjects comparison. Accordingly, the final sample of 39 participants, with approximately balanced gender representation (19 females and 20 males), remained adequately powered for the study objectives.

Second, we expanded the qualitative and behavioral analyses beyond the original protocol by incorporating observations from participant survey discussions and chat interactions with Windsurf. Since these chat-based interactions provided meaningful insights into participant behaviors, reasoning processes, and AI usage patterns, these were incorporated in both the Results, Section \ref{sec:behavioral}, and the Discussion, under Section \ref{sec:cgt_internet_comments}, to improve the comprehensiveness of the study.

Third, the originally planned ``suggestion modification rate'' metric was refined to measure \textit{immediate} modifications made to AI-generated suggestions rather than later downstream edits. We determined that immediate modifications provided a more reliable and interpretable measure of participant review and interaction behaviors directly following code generation events.

Finally, we introduced additional interaction-based behavioral metrics that were not part of the original preregistered analysis plan, but were part of CodeWatcher's captured data, including counts of AI and Internet related copy and paste actions, and chat interaction frequencies. These metrics were added to better characterize participant interaction patterns and sources of code usage throughout the study. These additions provide deeper insight into how participants interacted with CGTs.

\section{Results}

This section reports our results across three interrelated components of the study: (1) cognitive load as the primary psychological outcome (RQ1), (2) task performance as the behavioral outcome (RQ2), and (3) CGT usage behavior as a mediating interaction layer (RQ3). Cognitive load is further analyzed both as an outcome (RQ1) and as a predictor of performance outcomes (RQ2), allowing us to examine how tool usage influences cognitive processes and downstream task performance (RQ3).

\subsection{Participants}

%A total of 39 participants were recruited for this study. %To ensure fair competition in the coding tasks, participants with approximately one year of Python experience were required to confirm adequate Python proficiency. 
%The demographic and experience characteristics of the participants are summarized in Table \ref{tab:participants}.

%\begin{table}[ht!]
%\centering
%\caption{Participant Demographics and Background Characteristics}
%\label{tab:participants}
%\begin{tabular}{lcc}
%\toprule
% & \textbf{Count} & \textbf{Percentage (\%)} \\
%\midrule

%\textbf{Gender} & & \\
%Men & 20 & 53.8 \\
%Women & 19 & 46.2 \\
%\midrule

%\textbf{Age} & & \\
%18--24 & 18 & 46.2 \\
%25--34 & 21 & 53.8 \\
%\midrule

%\textbf{Education Level} & & \\
%Undergraduate (Year 4+) & 25 & 64.1 \\
%Graduate (Masters) & 8 & 20.5 \\
%PhD & 6 & 15.4 \\
%\midrule

%\textbf{Python Familiarity} & & \\
%$\sim$1 year & 7 & 17.9 \\
%2--3 years & 21 & 53.8 \\
%4+ years & 11 & 28.2 \\
%\midrule

%\textbf{CGT Usage Frequency} & & \\
%Always & 16 & 41.0 \\
%Most of the time & 13 & 33.3 \\
%Sometimes & 9 & 23.1 \\
%Rarely & 1 & 2.6 \\
%\midrule

%\textbf{Ethnicity} & & \\
%South Asian & 22 & 56.4 \\
%Asian & 7 & 17.9 \\
%African-American & 3 & 7.7 \\
%East Asian (China) & 2 & 5.1 \\
%Hispanic or Latino & 2 & 5.1 \\
%White/Caucasian & 2 & 5.1 \\
%West Asian (Iran) & 1 & 2.6 \\
%\bottomrule
%\end{tabular}
%\end{table}

\begin{figure}[t]
    \centering
    \includegraphics[width=1\linewidth]{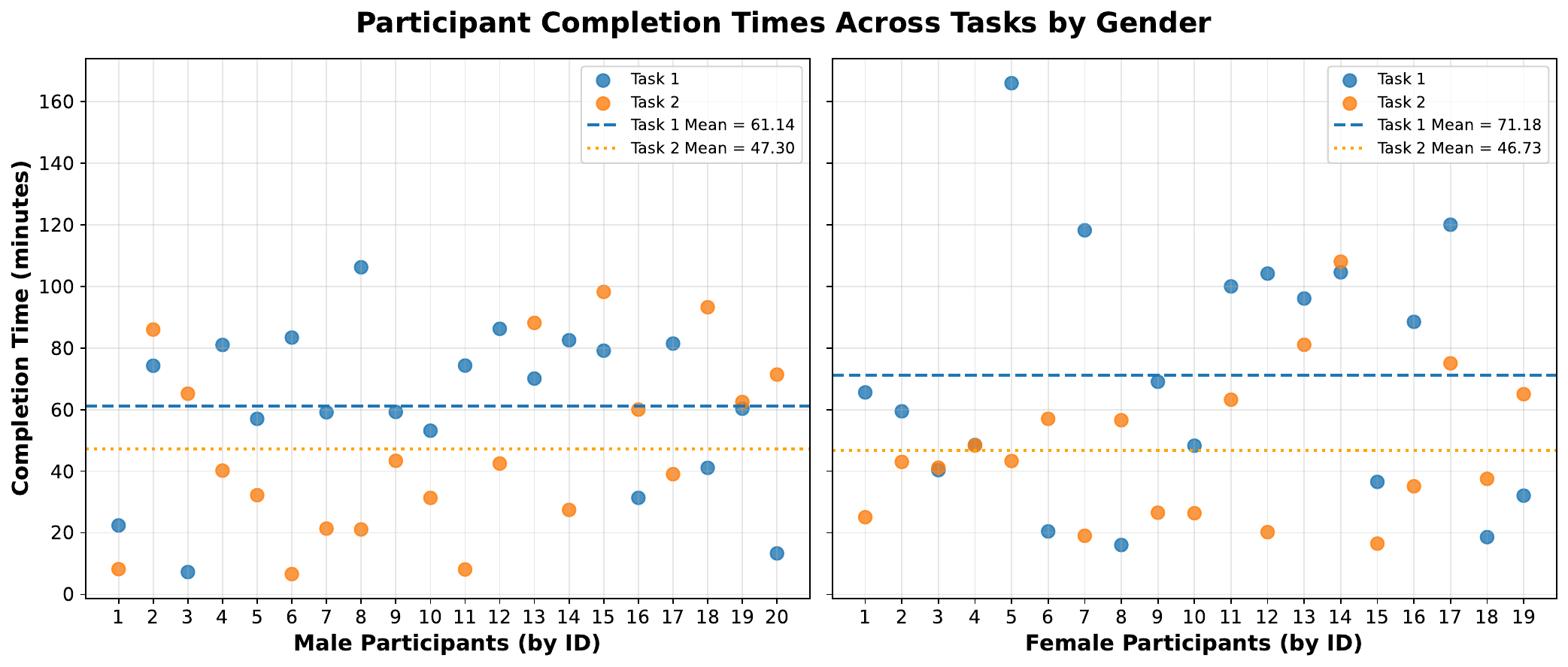}
    \caption{Participant completion times for Task 1 and Task 2.}
    \label{fig:completion_times}
\end{figure}

 A total of 39 participants were recruited for this study. Completion time exhibited considerable variation across individuals, ranging from 8.2 minutes to 97.5 minutes, indicating pronounced differences in task execution strategies. With a mean and standard deviation of 66.03 and 33.98 minutes for Task 1 and 47.02 and 26.39 minutes for Task 2 respectively as seen in Figure \ref{fig:completion_times}. The distribution was positively skewed, with a small number of slower participants increasing the overall mean relative to the median. In contrast, performance accuracy measures (core and advanced code correctness) were more tightly clustered, with several participants achieving near-ceiling scores, suggesting limited variance in correctness outcomes compared to efficiency-based measures. 

Outlier analysis further highlighted that variability was primarily driven by task completion time rather than code correctness, with only two participants identified as time outliers and none detected for performance metrics. This pattern suggests that participants adopted differing efficiency strategies while ultimately converging on similar correctness levels. Gender-based comparisons showed only small differences in both performance and cognitive load measures, with effect sizes indicating negligible-to-small practical differences. 

Overall, these findings suggest that while demographic groups were relatively balanced, individual differences in speed and interaction behaviour played a larger role than group-level effects in explaining variability in task performance.

\subsection[Cognitive Load and Gender (RQ1)]{RQ1: Are there significant gender differences in the cognitive load (intrinsic, extraneous, and germane load scores) experienced by participants when using CGTs to complete programming tasks?}

We analyzed differences in cognitive load across gender and tool condition using POM. Cognitive load was measured in three dimensions: intrinsic cognitive load (ICL), extraneous cognitive load (ECL), and germane cognitive load (GCL). Reliability analysis (Cronbach's alpha) indicated good internal consistency for all scales (ICL: $\alpha = 0.87$, ECL: $\alpha = 0.90$, GCL: $\alpha = 0.79$). Table~\ref{tab:cogload_performance} contains a summary of the results.

\subsubsection{Intrinsic Cognitive Load (ICL)}

\begin{table}[ht!]
\centering
\caption{Descriptive Statistics for ICL by Condition and Gender}
\begin{tabular}{l l c c c}
\hline
%Condition & Gender & $n$ & Mean ICL & SD \\
\textbf{Condition} & \textbf{Gender} & \textbf{$n$} & \textbf{Mean ICL} & \textbf{SD} \\
\hline
CGT       & Female & 19    & 2.59      & 1.45 \\
CGT       & Male   & 20    & 2.73      & 1.56 \\
Internet  & Female & 19    & 3.11      & 1.57 \\
Internet  & Male  & 20    & 3.65      & 1.92 \\
\hline
\end{tabular}
\label{tab:ds_icl}
\end{table}

As seen in Table~\ref{tab:ds_icl}, although ICL was comparable, female participants found the tasks easier than their male counterparts. POM results from the cumulative link mixed model indicated a significant effect of tool condition, with participants in the Internet condition reporting higher intrinsic cognitive load than those using CGT ($\beta = 1.01$, $SE = 0.45$, $z = 2.25$, $p = 0.025$). Gender effects were not significant ($\beta = 0.48$, $SE = 0.63$, $p = 0.45$), and Task effects were also non-significant. In other words, intrinsic cognitive load was primarily driven by tool usage rather than demographic differences, with the Internet condition being associated with higher perceived cognitive effort. That is, using the internet appears to be more tiring.

\subsubsection{Extraneous Cognitive Load (ECL)}

\begin{table}[ht!]
\centering
\caption{Descriptive Statistics for ECL by Condition and Gender}
\begin{tabular}{l l c c c}
\hline
%Condition & Gender & $n$ & Mean ECL & SD \\
\textbf{Condition} & \textbf{Gender} & \textbf{$n$} & \textbf{Mean ECL} & \textbf{SD} \\
\hline
CGT       & Female & 19    & 2.74      & 1.63 \\
CGT       & Male   & 20    & 3.03      & 2.09 \\
Internet  & Female & 19    & 3.79      & 1.91 \\
Internet  & Male   & 20    & 4.82      & 2.39 \\
\hline
\end{tabular}
\label{tab:ds_ecl}
\end{table}

For ECL, participant-level random effects could not be estimated due to negligible variance; therefore, a fixed-effects cumulative link model was used. Results indicated a significant effect of tool condition, with Internet participants reporting higher extraneous cognitive load ($\beta = 1.42$, $SE = 0.41$, $z = 3.43$, $p < 0.001$).

Task also had a significant effect ($\beta = -0.88$, $SE = 0.41$, $z = -2.17$, $p = 0.03$), suggesting slightly lower ECL on Task 2. Gender was not significant ($\beta = 0.56$, $SE = 0.40$, $p = 0.16$). In contrast to ICL, ECL was more strongly influenced by the interaction environment, with Internet-based task completion leading to higher potentially necessary cognitive effort, while gender did not meaningfully affect either load type. However, female participants had a lower ECL than the males as seen in Table~\ref{tab:ds_ecl}.
\subsubsection{Germane Cognitive Load (GCL)}

\begin{table}[ht!]
\centering
\caption{Descriptive Statistics for GCL by Condition and Gender}
\begin{tabular}{l l c c c}
\hline
%Condition & Gender & $n$ & Mean GCL & SD \\
\textbf{Condition} & \textbf{Gender} & \textbf{$n$} & \textbf{Mean GCL} & \textbf{SD} \\
\hline
CGT       & Female & 19    & 5.88      & 1.64 \\
CGT       & Male  & 20    & 5.62      & 1.92 \\
Internet  & Female & 19    & 6.19      & 1.86 \\
Internet  & Male   & 20    & 6.03      & 1.46 \\
\hline
\end{tabular}
\label{tab:ds_gcl}
\end{table}

The POM with random intercepts for participants indicated that GCL was generally higher in the Internet condition, suggesting that participants may have engaged in more effortful learning or mental processing. Although the effect did not reach statistical significance ($\beta = 0.67$, $SE = 0.42$, $z = 1.58$, $p = 0.11$). Gender and Task effects were not significant, yet unlike ICL and ECL, female participants demonstrated a higher GCL as seen in Table~\ref{tab:ds_gcl}.

\subsubsection{Cognitive Load and Task Performance}

\paragraph{Task Completion Time:} 
ECL predicted longer task completion times (Estimate = 3.87, $t \approx 2.06$), meaning the higher the ECL, the longer the task completion time. ICL and GCL were not significant predictors. Interaction models suggested that intrinsic load effects were slightly more pronounced in the Internet condition, whereas extraneous and germane effects were reduced when using CGTs. This indicates that higher extraneous load slows participants, but CGT usage may buffer the effects of intrinsic load on time.

\paragraph{Core Code Correctness} 
None of the cognitive load dimensions (ICL, ECL, GCL) significantly predicted core correctness ($p > 0.1$). Tool condition and gender also had no significant effect, suggesting that participants maintained basic task accuracy regardless of cognitive load or tool usage.

\paragraph{Advanced Code Correctness} 
For advanced correctness, \textbf{task 2 significantly improved outcomes} (Estimate = 0.86, $z \approx 2.49$, $p = 0.013$), while cognitive load measures and condition were not significant. This implies that advanced coding performance is more strongly influenced by task characteristics than cognitive load or tool choice.

\paragraph{Collinearity and Model Diagnostics} 
Variance inflation factors (VIFs) were low for all predictors (ICL = 1.51, ECL = 1.65, GCL = 1.05), indicating no problematic multicollinearity. Some singularity warnings were observed in core and advanced correctness models, but fixed-effect estimates remained stable.

\begin{table}[ht!]
\centering
\caption{Mixed-Effects Model Summary: Cognitive Load Predicting Task Outcomes}
\begin{tabular}{lcccc}
\hline
\textbf{Outcome} & \textbf{Predictor} & \textbf{Estimate} & \textbf{Std. Error} & \textbf{t / z} \\
\hline
Time Completion & ICL & 2.03 & 2.48 & 0.82 \\
Time Completion & ECL & 3.87 & 1.88 & 2.06 \\
Time Completion& GCL & 0.996 & 2.09 & 0.48 \\
Time Completion& Condition (Internet) & 4.78 & 6.48 & 0.74 \\
Time Completion& Task 2 & -10.57 & 6.13 & -1.72 \\
Time Completion& Gender (Male) & -3.91 & 7.67 & -0.51 \\
Core Correctness & ICL & -0.18 & 0.12 & -1.48 \\
Core Correctness & ECL & -0.08 & 0.10 & -0.79 \\
Core Correctness & GCL & -0.14 & 0.11 & -1.26 \\
Core Correctness & Task 2 & -0.26 & 0.36 & -0.72 \\
Adv Correctness & Task 2 & 0.86 & 0.35 & 2.49* \\
\hline
\multicolumn{5}{l}{* $p < 0.05$, ns = not significant}
\end{tabular}
\label{tab:cogload_performance}
\end{table}

\paragraph{Key Insights:} 
Task completion time is sensitive to extraneous load, suggesting that participants slow down under distracting or poorly designed interfaces for CGTs. Core correctness is robust across load dimensions, while advanced correctness depends primarily on task characteristics. CGT usage may help mitigate the effects of intrinsic load on completion time without directly improving correctness outcomes.

\begin{tcolorbox}[
    colback=blue!5!white, 
    colframe=blue!75!black, 
    title=\subsubsection*{RQ1 Summary},
     arc=2mm,
    boxrule=0.5mm
]
Overall, tool condition significantly influenced intrinsic and extraneous cognitive load, with participants using CGTs reporting lower ICL and ECL than those using Internet resources. Germane cognitive load was slightly higher in the Internet condition, though not statistically significant. Gender effects were generally small and non-significant across all cognitive measures. No statistically significant gender differences were observed across cognitive load relative to gender, providing no support for H1. %These results suggest that tool choice meaningfully affects cognitive load during programming tasks, supporting the potential cognitive benefits of CGT usage.
\end{tcolorbox}

\subsection[CGT Outcomes and Gender (RQ2)]{RQ2: How do CGT outcomes differ between participants of different genders?}

Task outcomes are evaluated using correctness metrics and completion time. %For both RQ2 and RQ3, a summary of results can be found in \textcolor{red}{Table \ref{tab:task_level_models}}. 

\subsubsection{Descriptive Statistics}

Descriptive statistics were calculated for core code correctness, advanced code correctness, and task completion time, grouped by tool condition, gender, and task. Means, standard deviations, and observed ranges are shown in Table~\ref{tab:descriptive_rq2}.

\begin{table}[htbp]
\centering
\caption{Descriptive Statistics for Task Outcomes by Condition, Gender, and Task (Means as Ratios)}
\label{tab:descriptive_rq2}
\begin{tabular}{l l l l r r}
\toprule
%Outcome & Condition & Gender & Task & Mean (Ratio) & SD \\
\textbf{Outcome} & \textbf{Condition} & \textbf{Gender} & \textbf{Task} & \textbf{Mean (Ratio)} & \textbf{SD} \\
\midrule
Core CC & CGT      & Female & Task 1 & 0.571 & 0.535 \\
        & CGT      & Female & Task 2 & 0.583 & 0.359 \\
        & CGT      & Male   & Task 1 & 0.800 & 0.422 \\
        & CGT      & Male   & Task 2 & 0.425 & 0.457 \\
        & Internet & Female & Task 1 & 0.333 & 0.492 \\
        & Internet & Female & Task 2 & 0.357 & 0.453 \\
        & Internet & Male   & Task 1 & 0.400 & 0.516 \\
        & Internet & Male   & Task 2 & 0.575 & 0.501 \\
\midrule
Advanced CC & CGT      & Female & Task 1 & 0.114 & 0.107 \\
            & CGT      & Female & Task 2 & 0.563 & 0.356 \\
            & CGT      & Male   & Task 1 & 0.160 & 0.084 \\
            & CGT      & Male   & Task 2 & 0.425 & 0.457 \\
            & Internet & Female & Task 1 & 0.067 & 0.098 \\
            & Internet & Female & Task 2 & 0.357 & 0.453 \\
            & Internet & Male   & Task 1 & 0.080 & 0.103 \\
            & Internet & Male   & Task 2 & 0.575 & 0.501 \\
\midrule
Completion (min) & CGT      & Female & Task 1 & 45.59 & 21.15 \\
                 & CGT      & Female & Task 2 & 51.03 & 28.19 \\
                 & CGT      & Male   & Task 1 & 61.06 & 25.05 \\
                 & CGT      & Male   & Task 2 & 40.24 & 33.32 \\
                 & Internet & Female & Task 1 & 86.10 & 43.58 \\
                 & Internet & Female & Task 2 & 39.37 & 13.92 \\
                 & Internet & Male   & Task 1 & 61.21 & 29.05 \\
                 & Internet & Male   & Task 2 & 54.36 & 24.25 \\
\bottomrule
\end{tabular}
\end{table}
\subsubsection{Code Correctness}

\paragraph{\textbf{Core Code Correctness.}}
A beta mixed-effects regression was used to model core code correctness. There was no statistically significant effect of tool condition ($\beta = -0.128$, $p = 0.654$), indicating that participants achieved comparable correctness when using CGTs and the Internet. A significant effect of task was observed ($\beta = 0.825$, $p = 0.005$), suggesting that correctness differed across tasks, with Task 2 yielding higher correctness overall. No significant gender differences were found ($\beta = 0.183$, $p = 0.517$).

Descriptive statistics (Table~\ref{tab:descriptive_rq2}) show that for Task 1, CGT users achieved higher mean correctness than Internet users across both genders (e.g., Male: $M=0.80$ vs. $M=0.40$). However, for Task 2, correctness levels were more comparable across conditions.

\paragraph{\textbf{Advanced Code Correctness.}}
For advanced correctness, the beta regression with participant-level random effects revealed no statistically significant effect of tool condition ($\beta = -0.128$, $p = 0.654$), indicating that CGTs did not significantly outperform the Internet in supporting advanced-level correctness. However, a statistically significant effect of task was observed ($\beta = 0.825$, $p = 0.005$), with Task 2 yielding higher advanced correctness scores overall. Gender differences were not statistically significant ($\beta = 0.183$, $p = 0.517$).

Descriptively, CGTs showed higher advanced correctness for Task 1 across both genders (e.g., Female $M=0.11$ vs. $M=0.07$; Male: $M=0.16$ vs. $M=0.08$). For Task 2, performance was similar or slightly higher in some Internet conditions (e.g., Male: $M=0.575$ vs. $M=0.425$), contributing to the lack of a significant overall condition effect.

%\begin{tcolorbox}[
%    colback=blue!5!white, 
%    colframe=blue!75!black, 
%    title=\subsubsection*{RQ2: Code Correctness},
%     arc=2mm,
%    boxrule=0.5mm
%]
%Overall, there were no statistically significant differences in either core or advanced correctness between CGT and Internet conditions. Task 1 had a statistically significant effect on correctness, with Task 2 generally yielding higher performance, while we did not observed statistically significant gender differences. %These findings suggest that while CGTs may offer descriptive advantages in certain scenarios (e.g., simpler tasks), they do not consistently lead to statistically significant improvements in correctness compared to traditional Internet-based approaches.
%\end{tcolorbox}

\subsubsection{Task Completion Time}

\paragraph{Model Results:}

A linear mixed-effects model revealed no statistically significant effect of tool condition on completion time ($\beta = 11.08$, $p > 0.05$), although tasks performed using the Internet took approximately 11 minutes longer on average. Task 2 was completed significantly faster than Task 1 ($\beta = -17.58$, $p < 0.05$), indicating substantial differences in task difficulty or familiarity. Gender differences were not statistically significant ($\beta = -4.74$, $p > 0.05$).

\paragraph{Key Insights:}
These results suggest that while CGTs may offer efficiency benefits, the difference in completion time is not statistically robust. Task characteristics play a much larger role in determining completion time than tool condition or gender. Task-related effects were more substantial than gender effects for performance outcomes. Task 2 was associated with faster completion times ($\beta = -17.58$) and significantly higher advanced correctness (OR $= 2.28$), whereas core correctness effects across condition and gender remained small and non-significant.

\begin{tcolorbox}[
    colback=blue!5!white, 
    colframe=blue!75!black, 
    title=\subsubsection*{RQ2: Summary},
     arc=2mm,
    boxrule=0.5mm
]
CGTs provide a measurable advantage in advanced correctness but do not significantly reduce completion time. Task characteristics strongly influence both correctness and efficiency, while gender does not significantly impact completion time or code correctness. Overall, there were no statistically significant differences in either core or advanced correctness between CGT and Internet conditions. The Task had a statistically significant effect on correctness, with Task 2 generally yielding higher performance, while we did not observed statistically significant gender differences. No statistically significant gender differences were observed across outcome metrics, providing no support for H2.
\end{tcolorbox}

\subsection[CGT Utilization and Gender (RQ3)]{RQ3: How does CGT utilization differ between participants of different genders?}

%We examine how participants utilize CGTs through several behavioral metrics.

\subsubsection{Descriptive Statistics}

Descriptive statistics for participant-level CGT utilization metrics are summarized in Table~\ref{tab:descriptive_rq3}. Metrics include suggestion acceptance rate (SAR), AI-generated code ratio, AI-modified code ratio, and prompt generation count, split by gender.

\begin{table}[htbp]
\centering
\caption{Descriptive Statistics for Participant-Level CGT Utilization Metrics by Gender}
\label{tab:descriptive_rq3}
\begin{tabular}{l l c c c c}
\toprule
%Metric & Gender & $n$ & Mean & SD & Range \\
\textbf{Metric} & \textbf{Gender} & \textbf{$n$} & \textbf{Mean} & \textbf{SD} & \textbf{Range} \\
\midrule
SAR Ratio           & Female & 19 & 0.98 & 0.32 & 0.15 – 0.4 \\
                    & Male   & 20 & 0.97 & 0.35 & 0.16 – 0.3 \\
AI Generated Ratio   & Female & 19 & 0.618  & 0.132  & 0.45 – 0.85 \\
                    & Male   & 20 & 0.607  & 0.125  & 0.46 – 0.82 \\
AI Modified Ratio    & Female & 19 & 0.045  & 0.019  & 0.02 – 0.08 \\
                    & Male   & 20 & 0.043  & 0.018  & 0.01 – 0.08 \\
Prompt Count         & Female & 19 & 12.3   & 4.5    & 6 – 21 \\
                    & Male   & 20 & 12.0   & 4.8    & 5 – 20 \\
\bottomrule
\end{tabular}
\end{table}

Effect sizes across AI usage and interaction behaviours were generally small, with most IRRs remaining close to 1.00, suggesting minimal practical differences across conditions or genders. The primary exception was AI insertions, where males demonstrated substantially higher insertion counts (IRR $\approx 6.16$).

\textbf{Suggestion Acceptance Rate (SAR):} 
A linear regression model revealed no statistically significant effect of gender on SAR ($\beta = -0.000035$, $p = 0.799$). The overall SAR was high across participants (mean $\approx 98\%$), indicating that participants accepted nearly all generated suggestions.

\paragraph{Key Insights:}
This suggests that users are not selective when incorporating CGT outputs, potentially reflecting a lack of critical evaluation of generated code. Participants were not selective and still utilized generated code after iterating through their options, this was confirmed in the participant recordings.

\textbf{CGT Generated Code Ratio:} 
The proportion of final code generated by CGTs was relatively high (mean $\approx 0.612 $ or 61.2\%). However, no statistically significant gender differences were observed ($\beta = -0.094$, $p = 0.405$).

\paragraph{Key Insights:}
Participants relied heavily on CGTs for code generation, regardless of gender. 

\textbf{CGT Code Modification Rate:}
The proportion of generated code that was subsequently modified was low (mean $\approx 0.044$ or 4.4\%). No statistically significant gender effect was found ($\beta = 0.0088$, $p = 0.535$).

\paragraph{Key Insights:}
This suggests that once accepted, CGT-generated code is rarely modified, indicating either trust in the output or limited verification.

\textbf{Prompt Generation Count:} 
A negative binomial model showed no statistically significant gender differences in prompt generation ($\beta = -0.068$, $p = 0.856$).

% \paragraph{Interpretation.}
% Participants generated a similar number of prompts regardless of gender, suggesting comparable interaction patterns with CGTs.

\paragraph{Key Insights:} In general, the previous results suggest that the way Males and Females used CGTs are not statistically different in our sample. 

\subsubsection{Correlation and Exploratory Analysis}
\label{sec:correlations}

To explore relationships between task performance, behavioral variables, and AI usage patterns, we computed Pearson correlation coefficients across all key study variables, including completion time, correctness measures, and interaction metrics (Table~\ref{tab:correlations}).

\begin{table*}[ht]
\centering
\caption{Pearson correlation matrix of task performance, interaction behavior, and AI usage variables.}
\label{tab:correlations}

\small
\begin{tabular}{lrrrrrrrr}
\toprule
%& C\_Time & Core & A\_C.C & Prompt & SAR & SM\_I & AI\_G. & AI\_M.\\
& \textbf{C\_Time} & \textbf{Core} & \textbf{A\_C.C} & \textbf{Prompt} & \textbf{SAR} & \textbf{SM\_I} & \textbf{AI\_G.} & \textbf{AI\_M.}\\
\midrule

\textbf{C\_Time}
& 1.000 & -0.027 & -0.217 & 0.276 & 0.076 & 0.016 & 0.120 & -0.128 \\

\textbf{Core C.C.}
& -0.027 & 1.000 & 0.672 & 0.122 & -0.031 & 0.020 & 0.020 & 0.021 \\

\textbf{A\_C.C.}
& -0.217 & 0.672 & 1.000 & 0.210 & 0.014 & -0.016 & 0.039 & 0.049 \\

\textbf{Prompt}
& 0.276 & 0.122 & 0.210 & 1.000 & 0.092 & -0.190 & 0.141 & -0.028 \\

\textbf{SAR}
& 0.076 & -0.031 & 0.014 & 0.092 & 1.000 & -0.027 & 0.277 & -0.168 \\

\textbf{SM\_I}
& 0.016 & 0.020 & -0.016 & -0.190 & -0.027 & 1.000 & -0.175 & 0.460 \\

\textbf{AI\_G}
& 0.120 & 0.020 & 0.039 & 0.141 & 0.277 & -0.175 & 1.000 & -0.353 \\

\textbf{AI\_M}
& -0.128 & 0.021 & 0.049 & -0.028 & -0.168 & 0.460 & -0.353 & 1.000 \\

\bottomrule
\multicolumn{9}{l}{C\_Time = Completion Time, C.C. = Code Correctness, A\_C.C. = Advanced C.C.} \\
\multicolumn{8}{l}{AI\_G = AI Generated Code, AI\_M = AI Modified Code} \\
\end{tabular}
\end{table*}

The results indicate a small but meaningful positive association between completion time and prompt usage ($r = 0.28$), suggesting that participants who issued more prompts tended to take longer to complete tasks. In contrast, completion time showed weak or negligible correlations with correctness measures for both core and advanced tasks ($r = -0.03$ and $r = -0.22$, respectively), indicating that faster completion was not strongly associated with reduced performance quality.

The two code correctness measures (core and advanced) were moderately to strongly correlated ($r = 0.67$), suggesting consistency in participant performance across task difficulty levels. This supports the internal validity of the correctness metrics as measuring a shared underlying construct of coding performance.

Regarding AI usage behavior, the AI-generated code ratio and AI-modified code ratio exhibited a moderate negative correlation ($r = -0.35$), indicating that participants who relied more heavily on generated code tended to modify it less frequently. Additionally, ratio of suggestions modified immediately (SM\_I) or within 30 seconds, showed a moderate positive association with AI-modified ratio ($r = 0.46$), suggesting that modification-related interaction behaviors are meaningfully aligned.

However, correlations between AI usage variables and performance outcomes (time and correctness) were generally weak ($\lvert r \rvert < 0.15$), suggesting limited direct linear relationships between the extent of AI usage and task performance.

Overall, the correlation analysis suggests that while AI interaction behaviors are internally consistent, they are only weakly associated with performance outcomes. Instead, interaction intensity (particularly prompting behavior) shows stronger relationships with completion time than raw AI usage metrics.

\paragraph{Effects of Age, CGT Experience, and Python Familiarity.}
We examined whether participant background variables, specifically age, CGT usage frequency, and Python familiarity can predict performance outcomes beyond task and condition effects in order to better explore hidden effects. Age was the only consistently statistically significant demographic predictor. Participants aged 25--34 demonstrated statistically significant and higher core code correctness compared to the reference group ($\beta = 0.977$, $p = 0.030$), although no statistically significant effects were observed for time or advanced code correctness. This may be tied to experience increasing with age. CGT usage exhibited a non-linear relationship with performance: infrequent users showed higher advanced correctness ($\beta = 1.905$, $p = 0.043$), while moderate usage showed a marginal negative trend ($\beta = -0.714$, $p = 0.086$), and heavy usage had no significant effect. Python familiarity did not significantly predict correctness or time outcomes, though it showed a weak positive association with prompt generation ($\beta = 0.883$, $p = 0.071$). Overall, these results suggest that age has a modest effect on basic code correctness, while CGT experience influences advanced performance in a non-linear manner, and prior programming experience does not strongly explain task performance in this setting.

\subsubsection{Model Fit and Explanatory Power}

To assess the explanatory power of the models, we report Akaike Information Criterion (AIC), marginal $R^2$ (variance explained by fixed effects), and conditional $R^2$ (variance explained by both fixed and random effects).

Across models, we observe a consistent pattern in which fixed effects explain a moderate proportion of variance, while participant-level random effects contribute additional explanatory power primarily in the time model.

For task completion time, the model achieved a conditional $R^2$ of 0.418 and a marginal $R^2$ of 0.226, indicating that approximately 22.6\% of the variance is explained by fixed effects (e.g., condition, task, cognitive load), while including participant-level differences increases explanatory power to 41.8\%. This suggests that individual differences in performance remain an important source of variability in time-based outcomes.

In contrast, code correctness models (core and advanced code correctness) show higher marginal explanatory power (0.341--0.383), but conditional $R^2$ values could not be reliably estimated due to near-zero random effect variance components. This indicates that participant-level random effects contribute minimally beyond fixed effects, suggesting weak hierarchical structure for correctness outcomes in this dataset.

AIC values further indicate relative model fit differences across outcomes. The core correctness model achieved the best relative fit (AIC = -141.3), followed by the advanced correctness model (AIC = -89.9). The time model showed substantially higher AIC values (AIC = 501.8), reflecting greater unexplained variance in continuous performance time compared to bounded correctness outcomes modeled using beta regression.

Overall, these results indicate that:
(i) time-based performance is moderately explained by both fixed and participant-level effects,
(ii) correctness outcomes are primarily driven by fixed effects with limited hierarchical structure, and
(iii) cognitive and behavioral predictors provide meaningful but incomplete explanations of performance variability.

\subsubsection{Behavioral Interaction Patterns and Cognitive Load}
\label{sec:behavioral}

To further understand how participants interact with CGTs during task completion, we analyzed fine-grained behavioral metrics captured through CodeWatcher, including copy/paste actions, and focus/unfocus transitions. These metrics provide insight into interaction strategies beyond aggregate usage measures to further validate our results.

%To more precisely characterize CGT utilization patterns examined in RQ3, we analyze fine-grained behavioral logs captured through CodeWatcher. These event-level traces operationalize the low-level interactions (e.g., copy/paste actions and focus/unfocus transitions) from which aggregate utilization metrics such as suggestion acceptance rate and CGT-generated code ratio are derived. This enables a more detailed decomposition of how utilization behaviors emerge during task execution, rather than treating them as solely aggregate outcomes.

\paragraph{Descriptive Behavioral Patterns.}

Table~\ref{tab:behavioral_descriptive} summarizes key behavioral metrics across condition, gender, and task. Overall, participants exhibited substantial interaction activity, with frequent copying, pasting, and AI insertions across both CGT and Internet conditions.

\begin{table}[ht!]
\centering
\caption{Descriptive Behavioral Metrics by Condition, Gender, and Task}
\label{tab:behavioral_descriptive}
\begin{tabular}{l l l l c}
\toprule
\textbf{Condition} & \textbf{Gender} & \textbf{Task} & \textbf{Metric} & \textbf{Mean} \\
\midrule
CGT & Female & Task 1 & AI Copies & 9.00 \\
CGT & Female & Task 2 & AI Copies & 29.10 \\
CGT & Male   & Task 1 & AI Copies & 9.25 \\
CGT & Male   & Task 2 & AI Copies & 26.00 \\
\midrule
CGT & Female & Task 1 & Chat Pastes & 15.60 \\
CGT & Female & Task 2 & Chat Pastes & 18.40 \\
CGT & Male   & Task 1 & Chat Pastes & 9.50 \\
CGT & Male   & Task 2 & Chat Pastes & 12.30 \\
\midrule
Internet & Female & Task 1 & Internet Copies & 29.10 \\
Internet & Female & Task 2 & Internet Copies & 9.00 \\
Internet & Male   & Task 1 & Internet Copies & 29.80 \\
Internet & Male   & Task 2 & Internet Copies & 11.40 \\
\bottomrule
\end{tabular}
\end{table}

These results indicate that interaction intensity varies more by \textbf{task} than by condition. For example, AI copy actions from the chat increased substantially in Task 2 for CGT users (e.g., Female: $M=29.10$ vs. $M=9.00$ in Task 1), suggesting more iterative refinement or reliance on generated content in more complex tasks.

\paragraph{\textbf{Behavioral Effects on Cognitive Load.}}

Linear mixed-effects models were used to examine whether interaction behaviors predict cognitive load. Results showed that neither focus nor unfocus counts significantly predicted intrinsic, extraneous, or germane cognitive load ($|t| < 1.0$ across models). 

However, tool condition remained a significant predictor of cognitive load. Participants in the Internet condition reported higher intrinsic load ($\beta = 0.88$, $t = 2.63$) and extraneous load ($\beta = 1.42$, $t = 2.70$), consistent with earlier findings.

Correlation analysis revealed weak positive associations between interaction metrics and cognitive load (internet focus and ICL: $r = 0.28$; internet unfocus and ICL: $r = 0.27$), suggesting that increased interaction or context switching may be mildly associated with higher perceived effort, though not strongly predictive.

\paragraph{\textbf{Behavioral Effects on Task Performance.}}

We further examined whether behavioral metrics predict task performance. Results indicate that most interaction measures, including focus/unfocus and copying behaviors, were not significant predictors of correctness or completion time ($p > 0.1$).

The only significant behavioral predictor was \textbf{CGT insertions}, which positively predicted completion time ($\beta = 0.123$, $t = 3.54$). This suggests that increased insertion of AI-generated code is associated with longer task durations, potentially reflecting iterative refinement, debugging, or integration overhead.

No behavioral variables significantly predicted core or advanced correctness, reinforcing earlier findings that correctness is largely driven by task characteristics rather than interaction patterns.

\paragraph{\textbf{Gender Effects in Behavioral Metrics.}}

Across all behavioral models, no statistically significant gender differences were observed ($p > 0.1$). While descriptively, some differences emerged (e.g., Females exhibiting slightly higher chat paste counts).%, these differences were not statistically robust.

\begin{tcolorbox}[
    colback=blue!5!white, 
    colframe=blue!75!black, 
    title=\subsubsection*{RQ3: Summary},
     arc=2mm,
    boxrule=0.5mm
]
Participants rely heavily on CGTs for code generation but selectively accept suggestions and rarely modify accepted code. Descriptive statistics show that the behavior is consistent across genders. No statistically significant gender differences were observed across utilization metrics, providing no support for H3. Additionally, tool condition did not significantly influence SAR, providing no support for H4.
\end{tcolorbox}

% Overall, behavioral interaction patterns reveal that:
% (i) participants engage in frequent and iterative interactions with both CGTs and Internet resources,
% (ii) interaction intensity varies primarily with task complexity rather than tool condition,
% and (iii) behavioral metrics have limited predictive power for cognitive load and correctness outcomes.

% The only notable effect is that increased AI insertions are associated with longer completion times, suggesting that integrating generated code introduces additional cognitive or workflow overhead. These findings highlight that while interaction behaviors reflect different usage strategies, they do not directly translate into improved performance outcomes.

\section{Discussion}

This study examined gender differences in CGT usage across task outcomes and utilization behaviors, while also comparing CGTs to traditional internet-based workflows. Overall, our findings challenge simplified assumptions about gender disparities in programming, revealing instead a more nuanced picture: while outcomes are largely equivalent across genders, both interaction patterns and tool affordances expose subtle differences and implications for inclusive CGT design.

\subsection{Participant Experiences Across Conditions}
\label{sec:cgt_internet_comments}

To complement the quantitative findings, we analyzed participants' open-ended comments across both the CGT and Internet conditions. The responses reveal both shared and divergent experiences in how participants approached coding tasks, particularly in terms of efficiency, learning, trust, and debugging.

\textbf{Efficiency vs. Effort Trade-off.} Participants in the CGT condition consistently emphasized speed and reduced effort, often describing the tool as allowing them to focus on outcomes rather than process. For example, P07 noted, ``only thing I need to care was the correct output,'' while P28 admitted they would ``dump generated code solutions and run it.'' In contrast, in the Internet condition participants highlighted the effortful and time-consuming nature of manual search and implementation. P06 reflected that the process ``took a lot of time'' and involved ``trial and error code,'' while P08 stated that ``most of the time was actually spent on searching relevant information.'' This contrast highlights a key trade-off: CGTs may be perceived as streamlining development relative to traditional internet-based workflows, however, the behavioral differences were limited, indicating that users’ subjective perceptions of efficiency and support may exert a stronger influence on their experience than measurable interaction outcomes, except in the case of GCL.

Aligned with this experience, recent empirical evaluations suggest that the perceived efficiency of CGTs may be characterized by a significant perception-reality gap. As demonstrated by Afroz et al.~\cite{afroz2025developer} in their application of the SPACE (Satisfaction, Performance, Activity, Communication, Efficiency) framework to developer productivity, while frequent generative AI users perceive faster task completion, these gains are often ``spurious", representing a surface-level acceleration that is offset by increased cognitive load from output verification and elevated code review burdens. This phenomenon, characterized as the ``slot machine effect,''\cite{Papalini2025} occurs because developers tend to remember the instantaneous ``jackpots'' of generated code while discounting the cumulative minutes spent repeatedly prompting the tool or cleaning up subtle errors. Consequently, the effort saved during the initial generation phase resurfaces downstream as maintenance and verification labor, creating an illusion of hyper-productivity that does not strictly align with objective project velocity. 

\textbf{Learning and Understanding vs. Outcome-Oriented Behavior.} A novel distinction emerged in how participants perceived learning. Internet users frequently described deeper engagement and skill development, even when unsuccessful which is supported by the observed higher GCL. For instance, P06 noted, ``I did actively learn throughout the task, and have a better understanding,'' and P36 described the experience as ``very knowledgeable.'' In contrast, CGT users often reported minimal engagement with the underlying code, with P08 stating, ``I made minimum effort to really understand what it wrote and assumed it would be correct.'' This suggests that while CGTs improve efficiency, they may reduce opportunities for learning and conceptual understanding, encouraging a more outcome-oriented approach.

This difference was also reflected in how participants used external resources. Internet users primarily relied on documentation and example-based sources such as Python documentation, Stack Overflow, GeeksforGeeks, and Real Python. Rather than searching for specific error messages, participants often reframed their queries around how to perform a particular task and frequently rewrote portions of their solution based on retrieved examples or tutorials. This example-driven workflow suggests a more active learning process, where participants engaged with alternative implementations and incrementally refined their understanding.

This observed shift toward outcome-oriented behavior is supported by recent randomized controlled trials on AI-assisted programming. For instance, Shen and Tamkin~\cite{shen2026ai} found that developers using AI assistance scored 17\% lower on subsequent comprehension tests compared to a hand-coding control group. Notably, participants who increasingly relied on AI, progressing from asking clarifying questions to delegating full code generation, demonstrated the largest declines in understanding. This result is even more broadly applicable and in line with recent studies about cognitive debt \cite{kosmyna2025your} caused by using generative AI tools. This pattern closely mirrors our findings, where CGT users reported minimal engagement with underlying code, suggesting that reduced cognitive effort during task execution may hinder schema construction and long-term learning.

\textbf{Trust and Verification:} CGT participants often demonstrated implicit trust in generated outputs, sometimes without verification. P02 noted they did not check documentation because the code appeared correct at a first glance, while P12 relied entirely on the tool's testing suggestions. In contrast, Internet users expressed uncertainty and the need for validation. For example, P37 stated, ``I honestly am not too sure if my code is correct at all,'' reflecting lower confidence despite greater engagement. This suggests that CGTs may increase perceived confidence regardless of actual correctness, whereas Internet-based approaches promote caution and verification.

This tendency toward unverified acceptance is reflected in recent large-scale empirical analyses of AI-generated code. For example, Liu et al. \cite{liu2026debt} found that over 15\% of AI-generated commits introduce issues, with 24.2\% persisting into later revisions. These findings align with our observations, suggesting that when developers rely on seemingly correct outputs without verification, errors are more likely to propagate. Importantly, many of these issues are not immediate runtime failures but more subtle code quality problems (e.g., code smells), which are less likely to be detected during outcome-oriented workflows \cite{liu2026debt,schreiber2025security}. This suggests that while CGTs may improve short-term efficiency, they risk introducing latent technical debt that degrades long-term maintainability, effectively shifting effort from development to future debugging and refactoring.

\textbf{Debugging Experiences.} Debugging emerged as a challenge in both conditions, but in different ways. CGT users frequently criticized the tool's inability to debug effectively, noting failures to interpret error messages or suggesting irrelevant implementations. In contrast, Internet users described debugging as time-consuming but ultimately manageable through external resources. For instance, P28 noted that ``debugging [was] a bit difficult and time-consuming,'' but identified ``Reddit and StackOverflow'' as reliable sources. Another participant highlighted that such difficulties are ``typical with regular coding,'' suggesting familiarity with the process despite its challenges.

This difference reflects a broader diagnostic gap observed in AI-assisted development. Because CGT users do not incrementally construct the underlying logic, they may lack the mental models needed to isolate and resolve errors when failures occur. Consistent with this, Shen and  Tamkin~\cite{shen2026ai} found that the largest performance differences between AI-assisted and manual programmers occurred specifically in debugging tasks. When AI tools fail to resolve issues, users may resort to iterative prompting without fully understanding the problem, limiting effective diagnosis~\cite{pimenova2025good,shen2026ai}. Together, these findings suggest that while CGTs support code generation, they can hinder the development of diagnostic reasoning skills needed for debugging.

\textbf{The Role of External Resources and Context.} The availability of external information played a critical role in shaping user experiences. CGT participants expressed frustration when unable to access documentation or provide additional context, with P17 noting it was ``incredibly frustrating not being able to look up documentation.'' Conversely, Internet users relied heavily on external resources, even if inefficient. P06 participant described the process as ``tedious'' but necessary, while P17 emphasized adapting examples from documentation to build working solutions. This highlights a key distinction: CGTs centralize assistance but may limit context, whereas Internet use distributes effort across multiple sources.

Contextual limitations are a known bottleneck in LLM performance, particularly in enterprise environments. As noted by Becker et al.~\cite{becker2025measuring}, advanced coding assistants struggle to access the ``tacit knowledge'' or implicit repository context that experienced maintainers possess. Furthermore, a recent benchmark on iterative code generation under stepwise refinement (SR-Eval)~\cite{zhan2025sr} demonstrated that the best-performing models achieve only a 20\% completion rate on repository-level tasks, as they fail to maintain cross-file consistency and domain-specific constraints across multiple turns. The inability of CGTs to ingest the broader systemic architecture forces developers to act as manual context-bridges, which can be more cognitively demanding than writing the code manually.

\textbf{Tooling and Environment Friction.} CGT users frequently commented on differences in model and tool quality, noting that some models were ``frankly unusable'' and that alternative tools performed better. Internet users, however, were more affected by setup overhead and environment configuration. One participant explained they had to ``install IDE, and test environment,'' which consumed significant time as some participants primarily code in AI-based IDEs like Cursor\footnote{https://cursor.com/}. This suggests that while CGTs reduce setup friction, they introduce variability in performance depending on the model and interface.

\textbf{Perceived Authenticity of Coding.} A particular theme emerged in the Internet condition, where P15 reflected that the experience ``took me back to pre 2022 (before GPT era) when coding felt more fulfilling.'' This sentiment was absent in the CGT condition and suggests that traditional workflows may be perceived as more authentic or rewarding despite being slightly less efficient.

This loss of fulfillment contrasts with the emerging paradigm of ``vibe coding,'' popularized by Karpathy~\cite{Karpathy2025VibeCoding}, where developers interact with AI primarily through natural language and accept generated code with minimal inspection. Others argue that this approach enhances flow and creativity by reducing low-level implementation effort and repositioning the developer as a high-level co-creator~\cite{pimenova2025good}.

However, our findings suggest that for many participants, fulfillment remains closely tied to the hands-on process of constructing code. As AI-assisted workflows remove this engagement, they may also diminish the sense of craftsmanship and ownership. Prior work similarly indicates that while generative tools can streamline workflows, they may reduce perceived authenticity and engagement \cite{verma2021artificial}. Together, these results suggest that shifting toward fully AI-mediated development may come at the cost of reduced satisfaction and a weakened sense of professional identity.

\textbf{Overall Comparison.} Across both conditions, participants recognized complementary strengths and weaknesses. CGTs were widely perceived as accelerating development and reducing cognitive load, which is consistent with our quantitative findings showing statistically significant lower ICL and ECL in the CGT condition. This reduction in cognitive burden encouraged a more outcome-oriented workflow, where participants focused on obtaining correct outputs rather than engaging deeply with the underlying code. However, this perceived efficiency did not translate into statistically significant improvements in completion time or correctness, suggesting that gains from reduced cognitive load and rapid code generation may be offset by challenges such as debugging limitations, model inaccuracies, and over-reliance on generated outputs. 

In contrast, Internet-based approaches were experienced as more effortful and cognitively demanding, reflected in higher ICL and ECL, but also promoted deeper engagement with the task. Participants reported more deliberate problem-solving, greater awareness of uncertainty, and stronger learning outcomes, aligning with the slightly higher (though non-significant) germane cognitive load (GCL) observed in this condition. This contrast highlights a perception–reality gap: while CGTs reduce cognitive load and make development feel easier, they do not necessarily improve objective efficiency and may shift effort away from comprehension toward execution. Notably, P17 explicitly stated, ``if given the choice between only internet and only llm, I will go with the internet any day of the week,'' underscoring a preference for control and reliability over convenience in certain contexts.

\begin{tcolorbox}[
    colback=blue!5!white, 
    colframe=blue!75!black, 
    title=\subsubsection*{Key Insights:},
    arc=2mm,
    boxrule=0.5mm
]
Across conditions, we observe a clear perception--reality gap in productivity and learning. CGTs reduce perceived effort by enabling rapid, low-friction generation, but shift work toward verification, debugging breakdowns, and repeated correction cycles without measurable gains in correctness or completion time. In contrast, Internet-based workflows are more effortful but support incremental reasoning, external validation, and stronger conceptual learning. Overall, CGTs optimize perceived efficiency, whereas traditional workflows better support understanding, suggesting that reduced cognitive load primarily redistributes effort rather than improving performance.
\end{tcolorbox}

\subsection{Observed Interaction Behaviors}
\label{sec:observations}

We analyzed screen recordings and Windsurf Chats of participants completing the tasks to further determine if their were any unique behavioral interactions. These observations provide insight into how participants interacted with tools in practice, revealing distinct behavioral patterns that help explain differences in cognitive load, learning, and perceived efficiency.

\textbf{Emergent Conversational Interaction Patterns.}

Across participants, CGT use exhibited a set of emergent conversational interaction patterns that collectively suggest users developed informal but evolving mental models of the system as a semi-collaborative programming agent rather than a passive tool. Participants frequently issued high-level generation requests such as ``generate unit tests,'' ``generate full code with unit tests,'' or broad rewrite instructions, while also pasting entire code blocks and terminal outputs into the chat with minimal contextual framing. In several cases, participants attributed system-level capabilities beyond the model’s actual functionality, including expectations that it could execute code or diagnose runtime and environment-level issues, indicating an inflated perception of agency. This was reflected in behaviors such as requesting execution or package/environment fixes, despite the absence of direct runtime access. This aligns with Kashumi Madampe et al.~\cite{madampe2025we} showed that developers often overestimate the autonomy and competence of AI programming assistants before encountering their contextual and architectural limitations. Consistent with Yunze Xiao et al.~\cite{xiao2025humanizing}, our findings suggest that even experienced users may overestimate CGT competence because conversational fluency, authoritative language, and human-like interaction patterns encourage anthropomorphic projection, leading users to subconsciously infer deeper situational awareness and operational agency than the systems actually possess.

Prompting strategies varied substantially across participants, ranging from terse, search-engine-like queries to emotionally charged directive imperatives. Concise prompts such as P09’s “how to select cursor” reflected moments where participants treated the CGT as a passive retrieval tool, leveraging efficient keyword-based interaction to minimize cognitive effort and obtain predictable outcomes. In contrast, capitalized directives such as ``NOW CAN YOU FIX IT” (P25, P33), and ``ugghh. stick to the problem'' (P03) typically emerged following repeated interaction failures, signaling elevated cognitive load, frustration, and a breakdown in the perceived human–AI collaborative dynamic. Consistent with work on uncertainty and resistance to innovation by IAM Verpaalen~\cite{verpaalen2024resistance}, these interactions suggest that users increasingly resorted to controlling language when system behavior became unpredictable. However, while directive imperatives may communicate urgency in human-to-human settings, emotionally charged prompts often lacked the semantic specificity required for effective AI assistance, contributing to vague outputs, iterative failure loops, and escalating user frustration.

An over-reliance on outcome correction is severely detrimental to long-term skill development. When participants prioritize rapid repair cycles without engaging in explanatory reasoning, they trap the AI in a "question-answer adjacency pair" pattern~\cite{bajpai2024let}. This forces the model to leap to action, providing premature fixes or sub-optimal outcome corrections without sufficient context or systemic investigation. While a smaller subset of participants in the observational data (such as P06) occasionally requested explanations of errors, indicating some variation in the depth of engagement, the dominant trend involving a failure to engage in explanatory reasoning transforms the developer from an active architect into a passive reviewer of alien code.  Similarly, participant P06's question—"What do I do with all these informations?"—perfectly encapsulates the cognitive overload that occurs when an LLM operates without adequate human constraint. When a participant types "good job" (P36) to an AI coding assistant, they are subconsciously applying their social learning framework to the machine. The user believes, on an intuitive level, that providing positive reinforcement will "teach" the AI what behavior is desired, encouraging it to produce similar outputs in the future. Demonstrating a belief of CGTs being a socially responsive collaborator.

\textbf{Copy–Paste-Replace an Iterative Regeneration Workflow.} A dominant behavior in the CGT condition combined copy–paste with iterative regeneration. Participants frequently copied entire solutions from the chat interface, ran the program, and upon encountering errors, either fully deleted the previous code or requested a revised solution from the chat and pasted that over the previously inserted code rather than debugging incrementally. This generate-and-test approach allowed participants to offload problem-solving to the model, aligning with lower intrinsic and extraneous cognitive load observed in the CGT condition. This was observed without gender difference, that is both male and female participants demonstrated this \emph{tinkering} behavior \cite{burnett2010gender}. While efficient for code generation, repeated regeneration cycles sometimes offset speed gains and limited deeper engagement with the code.

In many cases, participants fully deleted previously generated code and replaced it with new outputs instead of debugging or editing existing implementations. This behavior aligns with qualitative reports of ``dump[ing] generated code solutions and run[ning] it,'' and helps explain the lower intrinsic and extraneous cognitive load observed in the CGT condition, as participants offloaded problem-solving to the model rather than engaging in step-by-step reasoning.

This behavior provides a potential explanation for why reduced cognitive load in the CGT condition did not translate into significantly faster completion times: while generation is fast, repeated regeneration cycles and unresolved debugging inefficiencies may offset these gains.

This behavior resembles a generate-and-test workflow, where solutions are repeatedly produced and evaluated until one works~\cite{melegati2026role}. In this case, participants relied on the CGT to generate new code and used program execution as the test, replacing previous attempts instead of refining them~\cite{pimenova2025good}. This helps explain why time savings were limited: although code generation is fast, repeatedly generating and testing full solutions introduces delays similar to the time spent on manual debugging.

\textbf{Limited Direct Code Modification.} When modifications did occur, they were typically superficial, such as adjusting print statements or performing minor refactoring. Participants rarely engaged in deeper structural changes to the generated code. Additionally, some participants partially reused snippets from generated outputs rather than integrating them systematically. This limited engagement with the code supports the finding that CGTs encourage outcome-oriented workflows and may reduce opportunities for deeper comprehension, consistent with lower cognitive load but also reduced learning engagement.

This pattern is consistent with findings from \cite{cynthia2026we}, which show that agent-generated pull requests are less likely to be modified by developers before merging. Together, these results suggest that CGTs encourage acceptance over adaptation. By limiting code modification, developers may bypass deeper engagement with the logic, reducing opportunities to build a coherent mental model of the system. While this may lower effort during task execution, it also constrains learning and deeper understanding.

\textbf{Preference for Chat over Code Completion.} Participants showed a clear preference for chat-based interactions over inline code completion features. In many instances, participants ignored code completion entirely, instead relying on the chat to generate full solutions or revisions. This suggests that conversational interfaces may play a more central role in shaping user behavior than traditional IDE-based assistance, reinforcing the shift toward prompt-driven development observed in the qualitative data.

\textbf{Manual Reproduction of Generated Code.} Several participants manually typed out code from the chat rather than copying and pasting it directly. While relatively infrequent (approximately five observed instances), this behavior may indicate attempts to better understand the generated solution or adapt it incrementally. By converting the passive act of reading into the active act of typing, these developers artificially re-introduced the friction necessary for schema construction. In the context of the ICAP hypothesis (Interactive, Constructive, Active, Passive) established by \cite{chi2014icap}, typing out the code moves the learner from a purely passive reception state into an active psychomotor state, marginally increasing the likelihood of knowledge retention. This contrasts with the dominant copy–paste pattern and may reflect higher engagement or effort to internalize the code.

\textbf{Challenges with Environment and Libraries.} In both conditions, but particularly in Task 2, participants struggled with setting up required libraries and environments. In some cases, participants were unable to successfully run or test their code, leading them to proceed without verification. This observation complements participant-reported frustrations with missing context or documentation and highlights the importance of environment-related barriers, which are not fully mitigated by CGTs.

\textbf{Implications for Understanding Tool Use.} Overall, these observations suggest that CGTs fundamentally shift programming behavior from incremental development and debugging toward a generate-and-test paradigm. While this reduces cognitive effort during code production, it may also discourage deeper engagement with code and introduce inefficiencies during debugging. In contrast, the Internet condition more closely resembled traditional workflows, characterized by incremental edits, active debugging, and direct interaction with external resources. These behavioral differences provide a mechanistic explanation for the observed trade-offs between lower cognitive load, similar completion times, and reduced depth of understanding in the CGT condition.

\begin{tcolorbox}[
    colback=blue!5!white, 
    colframe=blue!75!black, 
    title=\subsubsection*{Key Insights:},
    arc=2mm,
    boxrule=0.5mm
]
CGT use induces a hybrid human--agent workflow where developers oscillate between search tool, teammate, and executable oracle, producing inflated agency perceptions and socially grounded interaction. Crucially, programming shifts from incremental debugging to a dominant copy--paste--regenerate loop, where full solutions are repeatedly replaced and executed rather than refined, collapsing stepwise debugging into a generate--test cycle. Despite lower perceived cognitive load, this regeneration-centric workflow degrades efficiency and comprehension by reducing direct code modification, weakening IDE engagement, and accumulating iterative overhead from repeated full-solution generation and unresolved system constraints.
\end{tcolorbox}

\subsection{Cognitive Offloading Without Performance Gains}

A central finding of this study is that CGTs reduce intrinsic and extraneous cognitive load, yet do not lead to statistically significant improvements in completion time or overall correctness. This challenges a common assumption in both cognitive load theory and prior CGT literature, that reducing extraneous load should directly translate into improved performance outcomes by freeing resources for germane processing and learning \cite{sweller1988,takir2011effect}. Prior work in CGTs suggests that assistance tools can improve efficiency by minimizing unnecessary mental effort~\cite{Glassman2020}. However, our findings suggest that while CGTs successfully reduce perceived effort, they do not necessarily improve task efficiency. This divergence indicates that cognitive offloading may introduce new forms of effort, particularly in verifying, debugging, and iterating on generated outputs.

This result aligns with emerging work showing that AI-assisted programming does not always yield consistent productivity gains. For example,Mendes et a.\cite{Mendes2024} report that while developers perceive AI tools as helpful, objective improvements depend heavily on task complexity and user expertise. Our findings extend this by showing that even when cognitive load is reduced, performance gains may be offset by downstream interaction costs.

\begin{tcolorbox}[
    colback=blue!5!white, 
    colframe=blue!75!black, 
    title=\subsubsection*{Key Insights:},
    arc=2mm,
    boxrule=0.5mm
]
CGTs reduce cognitive load without improving correctness or completion time, as effort is shifted from generation to verification and debugging. This challenges the assumption that cognitive offloading directly yields performance gains, suggesting instead a redistribution of effort rather than net efficiency improvements.
\end{tcolorbox}

\subsection{Shift from Constructive to Generative Programming}

Our behavioral observations reveal a fundamental shift in programming workflows: from incremental, constructive development toward a generate-and-test paradigm. Participants frequently copied entire solutions, executed them, and regenerated outputs rather than debugging iteratively.

This pattern is consistent with recent qualitative studies of AI-assisted development. Barke et al.~\cite{barke2023grounded} describe similar behaviors, where developers rely on large language models for rapid code generation and treat outputs as disposable artifacts rather than incrementally refined solutions. Likewise, Jayagopal et al.~\cite{Chasins2022} highlight how CGTs reshape programming into a process of prompt formulation and output evaluation.

Our findings extend this literature by linking these behavioral shifts to cognitive load and performance outcomes. The generate-and-test workflow reduces the need for step-by-step reasoning, explaining the lower intrinsic and extraneous cognitive load observed in the CGT condition. However, it also introduces inefficiencies when outputs fail, helping explain why reduced cognitive effort does not translate into faster completion times.

\begin{tcolorbox}[
    colback=blue!5!white, 
    colframe=blue!75!black, 
    title=\subsubsection*{Key Insights:},
    arc=2mm,
    boxrule=0.5mm
]
With CGTs, programming shifts from incremental debugging to a generate--test loop, where full solutions are repeatedly regenerated rather than refined. This reduces cognitive load but introduces iterative inefficiencies that offset performance gains.
\end{tcolorbox}

\subsection{Web-Based Workflows: Higher Effort, Similar Outcomes}

Our results provide a more nuanced understanding of Internet-based workflows, showing that while they impose  higher cognitive load, they do not lead to much worse performance outcomes. Participants in the Internet condition reported significantly higher intrinsic and extraneous cognitive load reflecting the effortful nature of searching, interpreting, and integrating information from multiple sources. This is further supported by behavioral data, where Internet users exhibited high copy activity (e.g., $\sim$29 copies in Task 1), suggesting frequent context switching and information gathering.

Despite this increased effort, Internet users achieved comparable core and advanced correctness to CGT users ($p > 0.05$), and completion time differences were not statistically significant. This indicates that traditional web-based workflows, although more cognitively demanding, remain effective for achieving correct solutions. Notably, ECL was the only significant predictor of completion time ($\beta = 3.87$), suggesting that inefficiencies in web-based workflows primarily manifest as slower progress rather than reduced accuracy.

However, GCL was slightly higher in the Internet condition (though not significant), aligning with qualitative reports of deeper learning and engagement. This suggests that the additional effort required by web-based approaches may promote more active problem-solving and schema construction, even if it does not translate into immediate performance gains.

Together, these findings reinforce a key distinction: Internet-based workflows distribute effort across search, comprehension, and debugging, leading to higher cognitive load but sustained performance, whereas CGTs centralize and compress this effort, reducing perceived difficulty without significantly improving outcomes. This highlights that the “cost” of using websites is primarily cognitive and temporal, rather than reflected in correctness, and suggests that their value may lie in supporting deeper engagement rather than efficiency.

\begin{tcolorbox}[
    colback=blue!5!white, 
    colframe=blue!75!black, 
    title=\subsubsection*{Key Insights:},
    arc=2mm,
    boxrule=0.5mm
]
Web-based workflows impose higher cognitive load due to distributed search and integration effort, yet achieve comparable correctness and completion time to CGTs. This indicates that their cost is primarily cognitive rather than performance-related, with effort distributed across comprehension and debugging rather than eliminated.
\end{tcolorbox}

\subsection{Efficiency–Engagement Trade-off}

A key theme emerging from both quantitative and qualitative results is an efficiency–engagement trade-off. CGTs enable faster code generation and reduce effort, but they also appear to reduce engagement with the underlying problem-solving process.

Participants in the Internet condition reported greater learning and understanding, which aligns with the slightly higher (though non-significant) germane cognitive load observed in this condition. This finding is consistent with cognitive load theory, where germane load reflects productive mental effort associated with schema construction~\cite{sweller1988}. It also aligns with educational research suggesting that effortful problem-solving promotes deeper learning~\cite{wang2025}.

In contrast, CGT users often focused on obtaining correct outputs with minimal engagement, echoing concerns raised in prior work about “automation complacency” and reduced cognitive involvement in AI-assisted tasks~\cite{Parasuraman1997,Baxter2012}. Our study contributes to this discussion by empirically demonstrating this trade-off in programming contexts, showing that reduced effort may come at the cost of reduced learning opportunities.

\begin{tcolorbox}[
    colback=blue!5!white, 
    colframe=blue!75!black, 
    title=\subsubsection*{Key Insights:},
    arc=2mm,
    boxrule=0.5mm
]
CGTs improve efficiency but reduce engagement with problem-solving, while Internet workflows require more effort yet promote deeper learning and schema construction. This reveals an efficiency--engagement trade-off where reduced cognitive effort may come at the cost of reduced learning opportunities.
\end{tcolorbox}

\subsection{Over-Reliance and Limited Code Understanding}

Across all participants, we observe high reliance on CGTs, with over 60\% of final code generated by the tool and minimal modification of accepted suggestions. This pattern indicates a tendency toward over-reliance and limited critical evaluation.

This finding aligns with recent studies highlighting risks of blind trust in AI-generated code. Prather et al.~\cite{Becker2023} show that developers often accept generated code without thorough validation, while Jayagopal et al.~\cite{Chasins2022} note challenges in debugging AI-generated outputs. Our results extend this work by quantifying the extent of reliance and showing that it persists across users regardless of gender.

Importantly, this behavior appears to be driven by the interaction paradigm rather than individual differences. The combination of high generation rates and low modification suggests that CGTs encourage a workflow where understanding is secondary to execution. From a FATE perspective, this raises concerns about accountability and transparency, as developers may incorporate code they do not fully understand. Meanwhile, from a software engineering perspective, this raises concerns about technical debt being accumulated.

\begin{tcolorbox}[
    colback=blue!5!white, 
    colframe=blue!75!black, 
    title=\subsubsection*{Key Insights:},
    arc=2mm,
    boxrule=0.5mm
]
CGT use is characterized by strong over-reliance, with most code accepted with minimal modification, indicating limited critical evaluation and understanding. This suggests that the interaction paradigm prioritizes execution over comprehension, raising concerns about accountability and accumulated technical debt.
\end{tcolorbox}

\subsection{Task Sensitivity and Context-Dependent Benefits}

Our results show that task characteristics play a dominant role in determining outcomes. While CGTs provide descriptive advantages in advanced correctness, these effects are not statistically robust across all tasks, and completion time is primarily driven by task differences rather than tool condition.

This finding aligns with prior work emphasizing the context-dependent nature of AI tool effectiveness. Mendes et al.~\cite{Mendes2024} and Zhang et al.~\cite{Glassman2020} both note that the benefits of intelligent tools are more pronounced in complex or reasoning-intensive tasks. Our results support this view, showing that CGTs may selectively support higher-level reasoning without uniformly improving performance.

At the same time, the lack of consistent gains challenges industry narratives that position CGTs as general-purpose productivity enhancers. Instead, our findings suggest that their value lies in specific contexts, particularly where abstraction and problem decomposition are required.

\begin{tcolorbox}[
    colback=blue!5!white, 
    colframe=blue!75!black, 
    title=\subsubsection*{Key Insights:},
    arc=2mm,
    boxrule=0.5mm
]
CGT effectiveness is highly task-dependent, with performance differences driven more by task complexity than tool use, challenging claims of universal productivity gains.
\end{tcolorbox}

\subsection{Fairness Beyond Outcomes: Gender and Interaction}

Consistent with prior work on gender and programming behavior~\cite{burnett2010gender,burnett2016gendermag}, we do not observe significant gender differences in task outcomes. This suggests that CGTs may act as performance-equalizing tools, enabling users with different backgrounds to achieve similar results.

However, our findings also highlight that fairness cannot be assessed solely through outcomes. Subtle differences in interaction patterns, combined with widespread over-reliance, point to the importance of process-based measures. 

Our study contributes to this perspective by showing that even in the absence of outcome disparities, CGTs reshape how users engage with programming tasks. This underscores the need for evaluation frameworks that consider both what users achieve and how they achieve it.

\begin{tcolorbox}[
    colback=blue!5!white, 
    colframe=blue!75!black, 
    title=\subsubsection*{Key Insights:},
    arc=2mm,
    boxrule=0.5mm
]
While CGTs appear to equalize performance outcomes across gender, fairness is revealed to be process-dependent, as interaction patterns and reliance behaviors differ despite similar end results.
\end{tcolorbox}

\subsection{Summary}

%Overall, our results do not support hypotheses H2–H4 regarding gender differences in outcomes and utilization or differences between CGT and internet workflows. However, we uncover subtle behavioral differences in how users interact with CGTs, alongside strong evidence of widespread reliance on generated code. These findings highlight that fairness in CGTs is not solely about equal outcomes, but also about accommodating diverse interaction patterns and promoting responsible use.
Across conditions, our results suggest that cognitive effort is redistributed rather than eliminated when moving from Internet-based workflows to CGT-mediated development. Internet use increased ICL, reflecting greater effort in understanding tasks, and ECL, reflecting the cost of interacting with external sources; however, only ECL significantly predicted completion time, indicating that delays are driven less by task comprehension itself and more by the overhead of moving outside the IDE. This finding implies that inefficiencies are not rooted in reasoning difficulty, but in interaction locality, suggesting potential value in non-agentic, in-IDE knowledge systems (e.g., documentation or Stack Overflow-like interfaces) that preserve contextual continuity without full task delegation to AI.

Despite CGTs reducing both intrinsic and extraneous cognitive load, these reductions did not translate into significant improvements in correctness or completion time, nor did overall AI usage predict performance outcomes. Instead, usage exhibited a non-linear relationship with performance: infrequent users achieved higher advanced correctness, while moderate usage was associated with weaker performance and heavy usage showed no clear advantage. This pattern suggests that effectiveness is shaped less by frequency of AI interaction and more by how AI is integrated into problem-solving workflows, potentially reflecting differences in metacognitive strategies, debugging practices, and selective verification. The fact that substantial variance in performance remains unexplained by measured cognitive and behavioral variables further reinforces that deeper individual differences, such as reasoning style, task decomposition strategy, and self-regulation, play a more central role than usage intensity alone.

Further, task effects and prior knowledge strongly shaped interaction behavior, with participants reporting Task 1 as more difficult due to reliance on early undergraduate material, while Task 2, despite higher CGT interaction intensity, was completed faster. This suggests that performance and interaction dynamics are not solely functions of complexity, but of familiarity and representational fit between task structure and CGT capabilities. More broadly, CGTs appear to shift users from acceptance to validation-based workflows, where conversational interfaces are preferred over code completion due to their higher perceived transparency and end-to-end resolution capability. However, this shift also introduces a different cognitive trade-off: while CGTs streamline production, they may weaken sustained engagement in construction and verification, altering how users evaluate correctness and increasing frustration when system outputs violate elevated performance expectations.

This is further reflected in cognitive load findings, where context switching behavior did not significantly predict ICL, ECL, or GCL, suggesting that observable interaction fragmentation is not a reliable proxy for cognitive strain. Instead, internet-based switching appears more closely tied to ICL, indicating strategic engagement for task understanding rather than overload. While switching increased completion time although not statistically significant, it did not significantly alter cognitive load, implying that temporal inefficiency is decoupled from subjective mental effort. Finally, participants demonstrated higher frustration when CGTs produced incorrect outputs compared to self-authored errors, reflecting not unconditional trust, but elevated performance expectations, reduced perceived control, and a shift in responsibility attribution during debugging.

The lack of statistically significant gender differences across CGT and Internet conditions in performance, completion time, and interaction measures suggests that, in this setting, CGTs do not produce clear performance gaps between genders despite prior work highlighting gendered differences in self-efficacy, risk-taking, and exploratory programming behaviours~\cite{burnett2016gendermag,durndell2002computer}. This challenges the assumption that these broader differences necessarily translate into measurable outcome disparities in modern AI-assisted development workflows, where higher-level abstraction may reduce the influence of such traits on final performance. However, this should not be interpreted as gender having no role; instead, any differences are likely to manifest in more subtle interaction patterns such as prompting style, verification strategies, or trust calibration, which are not fully captured by aggregate performance metrics. Overall, this suggests that inclusivity concerns in CGTs may be less about final outcomes and more about how different users engage with, interpret, and learn from AI-generated assistance.

Collectively, these findings suggest that CGTs do not simply improve or degrade performance, but fundamentally reconfigure where effort is allocated, away from search and synthesis and toward prompting, validation, and expectation management. This redistribution has implications beyond productivity: while AI may lower immediate cognitive barriers, it also reshapes engagement, self-efficacy, and the perceived value of effortful learning, with downstream consequences for how developers learn, debug, and derive satisfaction from software construction.

\section{Implications}

A central implication of our work is that CGTs should be understood as \emph{cognitive support tools} rather than purely productivity-enhancing technologies. While CGTs reduce intrinsic and extraneous cognitive load, these reductions do not consistently translate into improved correctness or efficiency. Instead, they reshape how developers approach programming tasks, shifting effort from code construction to output evaluation and iteration. This suggests that the value of CGTs lies not in uniformly accelerating development, but in supporting higher-level reasoning, abstraction, and rapid prototyping, particularly for complex tasks.

For developers, this implies the need for more deliberate and reflective use of CGTs. The observed combination of high AI-generated code usage and low modification rates indicates a tendency toward over-reliance and limited engagement with generated outputs. While such workflows may reduce effort in the short term, they introduce risks in debugging, maintenance, and long-term skill development. Developers may benefit from hybrid strategies that combine CGT-generated code with active verification and incremental refinement, ensuring that efficiency gains do not come at the cost of understanding.

For industry, our findings challenge prevailing assumptions that CGTs uniformly improve productivity. Although participants perceived CGTs as efficient and less effortful, objective improvements in completion time and correctness were not statistically robust. This highlights the importance of rethinking how productivity is measured and valued in AI-assisted development. Rather than focusing solely on speed or output, organizations should consider how CGTs support reasoning, reduce onboarding barriers, and enable exploratory development. At the same time, the risks of over-reliance and shallow engagement highlighting the need for tooling that promotes transparency, explainability, and debugging support, particularly in high-stakes or production environments.

For academia, the results raise important questions about learning and assessment in the presence of CGTs. The shift toward outcome-oriented workflows, where participants focus on obtaining correct outputs with minimal engagement, suggests that traditional evaluation metrics such as correctness and completion time may no longer capture meaningful differences in learning. The higher germane engagement observed in Internet-based workflows further indicates that effortful problem-solving remains critical for developing deep understanding. This suggests that educational settings should emphasize code comprehension, debugging, and critical evaluation, and carefully consider how and when CGTs are introduced into curricula.

From a fairness perspective, our findings demonstrate that equal outcomes across genders do not imply identical interaction experiences. Given performance was similar across CGT and Internet conditions, CGTs may simply shift the strategy of problem solving from search-and-synthesis to prompting-and-verification without substantially changing overall outcomes. This highlights the need to move beyond outcome-based evaluations of fairness toward a broader consideration of process, agency, and user experience. Evaluations of CGTs should therefore incorporate both performance and behavioral metrics, examining how different users engage with and are supported by these tools.

Finally, our results point to important directions for future research. The lack of significant differences in traditional performance metrics, despite clear behavioral and cognitive differences, suggests that current evaluation paradigms may be insufficient. Future work should investigate fine-grained interaction patterns, longitudinal effects on learning, and the integration of CGTs within broader development ecosystems. Additionally, expanding beyond gender to consider other dimensions of diversity, such as experience level and language background, will be critical for developing a more comprehensive understanding of inclusivity in AI-assisted programming.

\section{Conclusion and Future Work}

Our study examined the effects of code generation tools (CGTs) versus Internet resources on cognitive load and task performance in programming tasks. Results indicate that tool choice significantly influences cognitive load: participants using CGTs reported lower intrinsic and extraneous cognitive load compared to those relying on Internet resources. Task outcomes showed that CGTs were particularly beneficial for complex coding tasks, enhancing advanced code correctness without substantially affecting core correctness or completion time. Gender effects were generally small and non-significant, suggesting the observed benefits of CGTs are consistent across male and female participants. Overall, these findings highlight the potential cognitive and performance advantages of integrating CGTs into programming workflows.

Future research could extend this work in several directions. First, larger and more diverse participant samples would improve generalizability and allow examination of interactions with programming experience. Second, longitudinal studies could assess how prolonged CGT usage affects skill acquisition and cognitive load over time. Third, exploring additional metrics such as code quality, readability, and maintainability would provide a more comprehensive understanding of tool impact. Finally, incorporating adaptive CGT interfaces that respond to users' cognitive load in real-time may further enhance both learning and performance in programming tasks.

\section{Acknowledgements}

\section{Declarations}
\vspace{-1em}
The authors make the following declarations:

\vspace{-1.5em}

\subsection{Funding}
\vspace{-1em}
We thank GOOGLECA (PWPU GR028067) and CNPq (420406/2023-9, 444802/2024-0, 408817/ 2024-0 and 308101/2025-1) for funding this research.

\vspace{-1.5em}

\subsection{Ethical Approval}
\vspace{-1em}
Ethics certificate H25-00502.

\vspace{-1.5em}

\subsection{Informed Consent}
\vspace{-1em}
All human data or personally identifiable information collected or analyzed is stored privately, and any participant identifiable data will not be available in the replication package.

\vspace{-1.5em}

\subsection{Author Contributions}
\vspace{-1em}
\textbf{Manaal Basha} contributed to the conception and design of the study; co-designed and implemented RQ1 , RQ2, and RQ3 under the supervision of the Last Author; participated in all aspects of the study trials; contributed to manuscript writing and revision.\\
\\
\textbf{Ivan Beschastnikh} contributed to the conception and design of the study, and contributed to the writing and revision of the manuscript.\\
\\
\textbf{Cleidson R. B. de Souza} contributed to the conception and design of the study, and contributed to the writing and revision of the manuscript.\\
\\
\textbf{Gema Rodr\'iguez-P\'erez} contributed to the conception and design of the study; supervised RQ1, RQ2, and RQ3; co-designed and assisted with all aspect of the study; and contributed to writing and revision of the manuscript.\\
\\
All authors read and approved the final version of the manuscript.

\vspace{-1.5em}

\subsection{Data Availability Statement}
\vspace{-1em}
All scripts, datasets, and supplementary materials are publicly available in our replication package~\cite{inclusiveai}.\footnote{\url{https://osf.io/tcfjr}}

\vspace{-1.5em}

\subsection{Conflict of Interest}
\vspace{-1em}
The authors declare that they have no conflict of interest.

\vspace{-1.5em}

\subsection{Clinical Trial Number}
\vspace{-1em}
Not applicable.

\markboth{References}{References}

%\bibliographystyle{IEEEtran}
% argument is your BibTeX string definitions and bibliography database(s)
\bibliographystyle{spmpsci}      % mathematics and physical sciences
\bibliography{software}   % name your BibTeX data base

\end{document}